\documentclass[%
 aip,
 amsmath,amssymb,
 reprint,%
]{revtex4-2}

\usepackage{graphicx}
\usepackage{dcolumn}
\usepackage{bm}

\usepackage[utf8]{inputenc}
\usepackage[T1]{fontenc}
\usepackage{mathptmx}
\usepackage[dvipsnames]{xcolor}

\begin{document}

\preprint{AIP/123-QED}

\title[Creep and drainage in the fast destabilization of emulsions]{Creep and drainage in the fast destabilization of emulsions}

\author{Riande I. Dekker}
\email{r.i.dekker@uva.nl}
\affiliation{Van der Waals-Zeeman Institute, Institute of Physics, University of Amsterdam, Science Park 904, 1098 XH Amsterdam, The Netherlands.}
\affiliation{Van 't Hoff Laboratory of Physical and Colloid Chemistry, Debye Institute for Nanomaterials Science, Utrecht University, Padualaan 8, 3584 CH, Utrecht, The Netherlands.}
\author{Antoine Deblais}
\affiliation{Van der Waals-Zeeman Institute, Institute of Physics, University of Amsterdam, Science Park 904, 1098 XH Amsterdam, The Netherlands.}
\affiliation{Unilever Foods Innovation Center, Bronland 14, 6708 WH Wageningen, The Netherlands.}
\author{Bastiaan Veltkamp}
\affiliation{Van der Waals-Zeeman Institute, Institute of Physics, University of Amsterdam, Science Park 904, 1098 XH Amsterdam, The Netherlands.}
\author{Peter Veenstra}
\affiliation{Shell Global Solutions International B.V., Grasweg 31, 1031 HW Amsterdam, The Netherlands.}
\author{Willem K. Kegel}
\affiliation{Van 't Hoff Laboratory of Physical and Colloid Chemistry, Debye Institute for Nanomaterials Science, Utrecht University, Padualaan 8, 3584 CH, Utrecht, The Netherlands.}
\author{Daniel Bonn}
\affiliation{Van der Waals-Zeeman Institute, Institute of Physics, University of Amsterdam, Science Park 904, 1098 XH Amsterdam, The Netherlands.}

\date{\today}

\begin{abstract}
The destabilization of emulsions is important for many applications, but remains incompletely understood. We perform squeeze flow measurements on oil-in-water emulsions, finding that the spontaneous destabilization of emulsions is generally very slow under normal conditions, with a characteristic time scale given by the drainage of the continuous phase and the coalescence of the dispersed phase. We show that if the emulsion is compressed between two plates, the destabilization can be sped up significantly; on the one hand the drainage is faster due to the application of the squeezing force. On the other hand, creep processes lead to rearrangements that also contribute to the destabilization.

\end{abstract}

\maketitle

\section{Introduction}
Emulsions play an important role in everyday life and can be found in, for example, food, cosmetics and paint. They consist of two immiscible liquids where one liquid is dispersed in the other \cite{Leal-Calderon2007}. Emulsions are thermodynamically unstable but can be kinetically stabilized by addition of surface-active agents. Indeed, stabilizing or destabilizing emulsions is of considerable industrial importance \cite{Bibette1992,Wasan2004}. For instance, stabilization is needed to improve the shelf life of cosmetics, drugs and food products \cite{Bergenstahl1998,Glampedaki2014,Degrand2016} while destabilization is a key step in oil recovery, by extracting water from the recovered crude oil. However, present techniques to destabilize emulsions are often energy-intensive and/or use chemical additives that end up in the water phase \cite{White2000,Pena2005,Less2008,Katepalli2016}.

Improving oil recovery by finding better ways to destabilize crude oil emulsions is still an important research topic. However, most research is based on improving already existing techniques, instead of finding new ways to destabilize emulsions. In the last decade, research has been focusing mainly on the use of chemical additives. Various surfactants have been investigated for their ability to induce destabilization \cite{Djuve2001,Fan2009,Mirvakili2012,Jia2017,Kumari2019}, but the last years the focus has also shifted to using biosurfactants \cite{Nasiri2020,Samak2020,Yang2020} and surface active ionic liquids \cite{Sakthivel2017,Jia2018,Liu2019b}. However, the main issue remains that these chemical additives acquire additional steps to separate them from the water phase. Other lines of research try to improve the technique of low salinity water flooding \cite{Chen2017,Bidhendi2018,Takeya2019,Sanyal2019}. This technique alone is not enough to recover all the crude oil though. Research on mechanical or electrostatic destabilization of crude oil emulsions is much less extensive \cite{Less2008,IliaAnisa2011,Rodionova2014}, which leaves a way to new, chemical free techniques to destabilize emulsions.   

Many factors influence the stability of emulsions \cite{Bremond2011}. One important phenomenon that drives destabilization of emulsions is Ostwald ripening, a process that involves the diffusion of the dispersed phase through the continuous phase. This results in the growth of bigger droplets at the expense of smaller ones \cite{Leal-Calderon2007,Brailsford1979,Yao1992}. However, this process is usually very slow, especially when the solubility of the dispersed phase in the continuous phase is small. Coalescence, which is coarsening by droplets merging together, is another and faster destabilization factor \cite{Leal-Calderon2007}. If coalescence is fast enough, its propagation through a concentrated emulsion can even lead to  full phase inversion of the emulsion \cite{Bremond2011}. Nonetheless, what governs the speed of coalescence is not very clear. It was shown recently that rapid coalescence can be induced by the simple mechanical action of, for example, a rigid blade. The emulsion between the blade and a surface destabilizes in the close vicinity of the contact line and in this way a rapid phase separation is achieved \cite{Deblais2015}. 

The central idea behind the blade technique is that the small gap between the blade and the plate deforms the emulsion droplets until they coalesce. Based on this idea, recently a new technique was proposed to destabilize emulsions using a simple squeeze flow. This allows to investigate the destabilization mechanism of highly-concentrated surfactant-stabilized oil-in-water emulsions in detail\cite{Dekker2020}. By decreasing the sample thickness of the emulsion, the oil droplets deform and water is squeezed out, thereby creating very thin films between the droplets that eventually break. A first film rupture then leads to a cascade of coalescence events, resulting in the destabilization of the emulsion. While this  technique can be of great interest for the oil industry, scaling it up requires more insight in especially the characteristic time scales that play a role in the destabilization. 

In this article, we use the squeeze flow technique for emulsion destabilization as described in \cite{Dekker2020} to investigate the dynamics and hence the characteristic time scales of the processes responsible for the destabilization of the emulsion. We first show the importance of the boundary conditions by using smooth and rough walls, and show that wall slip avoids coalescence when using smooth surfaces. Second, characteristic relaxation times during the deformation and coalescence are investigated by measuring the time dependence of the squeezing force, and comparing them to the time scales for creep of the emulsion and Darcy flow of the continuous phase through the network of the dispersed phase. 

\section{Methods}
\subsection{Emulsion preparation}
Silicone oil-in-glycerol/water emulsions stabilized by sodium dodecyl sulfate (SDS) are used as model emulsions in the majority of the experiments described in this paper. The continuous phase is prepared by dissolving 1 wt\% of SDS (Sigma-Aldrich) in a 50:50 mixture of glycerol and demineralized water. Nile red (Sigma-Aldrich) is added to the silicone oil (VWR Chemicals, viscosity 500 cSt) as a dye. The silicone oil is then slowly added to the continuous phase while stirring with a Silverson L5M-A emulsifier at 6,000 rpm obtaining an 80 v\% silicone oil-in-glycerol/water emulsion with an average droplet diameter of 20 $\mu$m. Due to the addition of glycerol to the water, the continuous phase is index-matched with the silicone oil, making the emulsion transparent. The combination of the protein bovine serum albumin (BSA, Sigma-Aldrich) and co-surfactant propylene glycerol alginate (PGA, Dextra) is used to stabilize 80v\% silicone oil-in-water emulsions. The continuous phase is prepared by dissolving 0.4 wt\% BSA and 0.4 wt\% PGA in water. Rhodamine B (Sigma-Aldrich) is added as a dye to the continuous phase. The mixing speed is kept at 6,000 rpm, resulting in an average droplet diameter of 42 $\mu$m.

For the water drainage experiments a different emulsion of 40 v\% castor oil in demineralized water with 1 wt\% is used. The emulsion is prepared in the same way as described before using 80 v\% of oil and afterwards diluted with the continuous phase to obtain an emulsion with 40 v\% of oil. 

\subsection{Squeeze flow experiments}
The squeeze flow experiments follow the same protocol as described by Dekker \textit{et al.} \cite{Dekker2020} Imaging of the emulsions is done using a confocal laser scanning microscope (CLSM, Zeiss LSM 5 Pascal). Adding a fluorescent dye (Nile Red) that is soluble in the oil phase but insoluble in the continuous aqueous phase allows us to use fluorescence microscopy to distinguish between the dispersed and the continuous phases in the emulsion layer and follow the deformation of the emulsion droplets under confinement. A droplet of emulsion is placed on a 1 mm thick glass plate, strong enough to support mechanical stresses of a few Newton. In most of the experiments, the glass substrate is sandblasted to obtain a roughness with a grit designation of P800. This refers to a roughness of 22 $\mu$m, which is close to the average droplet diameter in our emulsions. The rough substrate is used to avoid surface slip of the oil droplets during the experiments. A smooth cover slide of 170 $\mu$m thickness is placed on top of the emulsion droplet, thereby spreading the emulsion over the glass substrate. This results in an emulsion layer of roughly 50 $\mu$m thickness. Nile red is excited at 488 nm with an Argon laser (LASOS). We use an air objective with a magnification of $40\times$ and a numerical aperture NA = 0.75 (Zeiss EC Plan-Neofluar) which allows us to observe the emulsion film with a depth of field (DOF) of maximum 2 $\mu$m. The objective is corrected for the 1 mm thick glass substrate. 

To perform the squeezing experiments, a rheometer head (Anton Paar DSR 301) is mounted on top of the inverted confocal microscope. A parallel-plate geometry with a diameter of 10 mm is used as an upper geometry. The rheometer is held on a controllable precision vertical translation stage, which allows us to manually alter the height of the geometry while measuring the squeezing force using the normal force transducer of the rheometer. A schematic overview of the setup is shown in \cite{Dekker2020}. The parallel-plate geometry is brought closer to the glass substrate in a step-wise manner, with a step size of 1 to 2 $\mu$m, thereby slowly squeezing the emulsion. The thickness of the emulsion layer is measured by scanning in the $z$-direction. The central height of the emulsion layer is used to image the progress of deformation and destabilization of the emulsion. 

\subsection{Squeeze force measurements}
To investigate the behavior of the squeezing force in a more controlled manner, squeezing experiments are performed using an Anton Paar MCR302 rheometer where the sample thickness can be controlled more accurately than in the squeeze flow experiments as the up and down motion is controlled by a step motor. The experimental setup is kept similar to the squeezing experiments described before. A smooth cover slide of 170 $\mu$m thickness (20 mm length and width) is glued at the end of a parallel-plate geometry with a diameter of 10 mm. A droplet of emulsion is placed on a rough (P800) substrate and the gap between the geometry and the substrate is reduced from 45 to 5 $\mu$m in steps of 5 $\mu$m. A waiting time of 2000 s between every change in gap size is applied to allow the system to relax and over which the normal force is recorded.

\subsection{Creep measurements}
For the creep measurements, an Anton Paar MCR302 rheometer is used with a 50 mm-diameter parallel-plate measuring system. The emulsion is deformed for a period of 20 s with a shear stress of 5 Pa. After this deformation, the emulsion is allowed to recover over a period of 40 s. The gap size between the plates is varied from 45 to 5 $\mu$m in steps of 5 $\mu$m. During the creep experiment, the strain is measured.

\section{Results and discussion}

\subsection{Water drainage in emulsions}
\begin{figure}
	\centering
	\includegraphics[width=\linewidth]{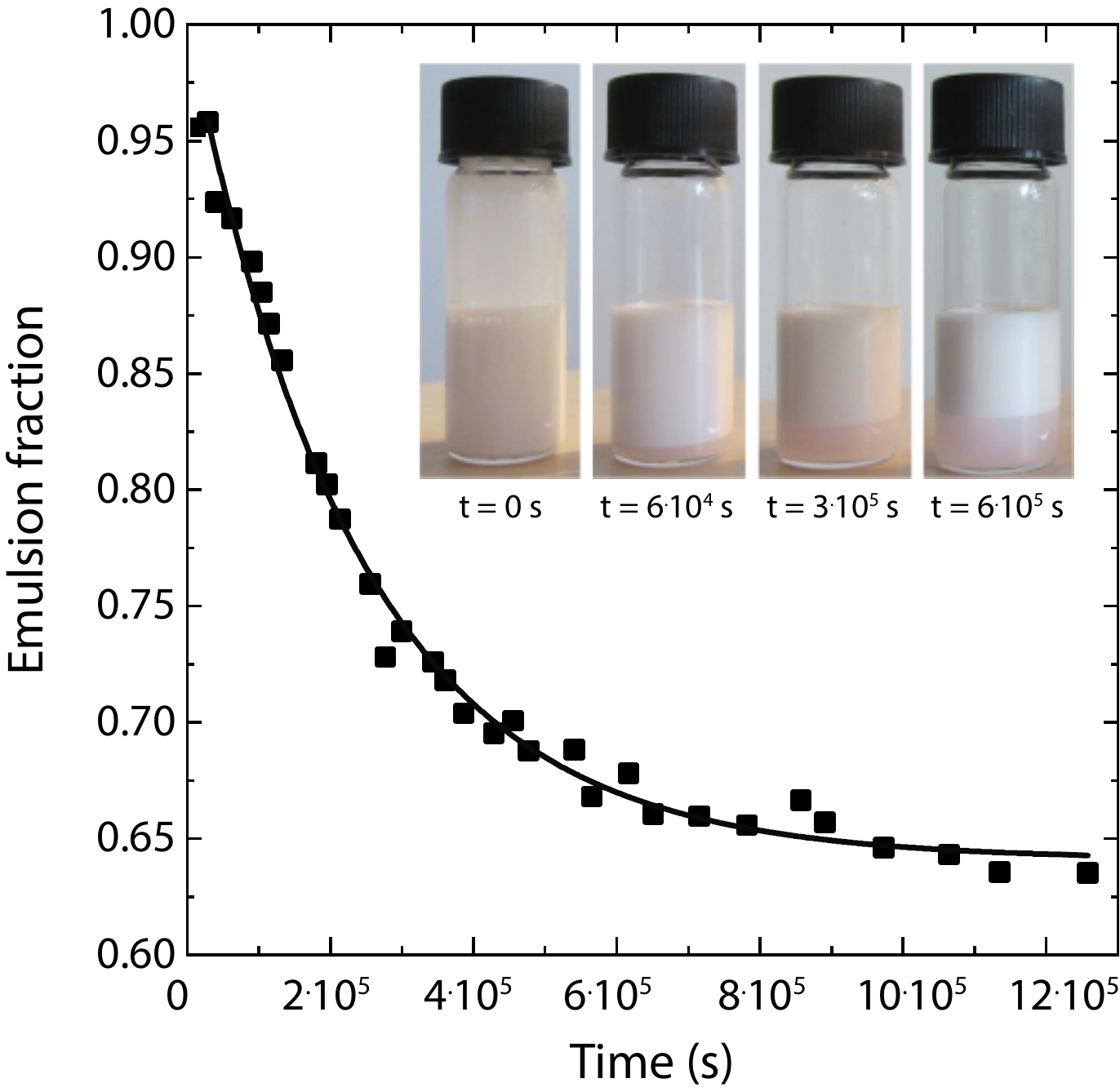}
	\caption{Drainage of water from a 40\% oil-in-water emulsion as a function of time, as calculated by measuring the emulsion height as a fraction of total liquid height. The emulsion has an oil fraction of below the jamming transition (64\%). The solid line is an exponential fit with a drainage time of $2.3 \cdot 10^6 $ s. Insets: photographs at four consecutive times.} 
	\label{hartland}
\end{figure}

Water drainage is a familiar phenomenon in oil-in-water emulsions and is most efficient at oil fractions below 64 \%. At this volume fraction, the transition to jamming occurs. We therefore investigate the water drainage in a 40 v\% castor-oil-in-water emulsion by measuring the heights of the emulsion layer and the emergent water layer as a function of time. As shown in Figure \ref{hartland}, the water drains to the bottom of the vial and consequently the fraction of the total height occupied by emulsion decreases over time. The photographs in the upper right of Figure \ref{hartland} reveal that, on the bottom of the vials, a pink layer of water emerges. The pink color is due to the presence of Rhodamine B. We do not observe a layer of oil forming on top of the emulsion, so no coalescence of the oil droplets is observed. Similar experiments performed on crude oil-in-water emulsions obtained during oil recovery \cite{Hartland1987,Aleem2020}, do show oil drop coalescence implying that our model emulsions are more stable to coalescence than the crude oil-in-water emulsions, and so more difficult to destabilize. 

The decrease of the emulsion volume due to the drainage is well fitted with a simple exponential (solid line in Figure \ref{hartland}) allowing to assess a characteristic time of the drainage and hence the partial emulsion destabilization. From the exponential fit a drainage time of $2.3 \cdot 10^6 $ s is obtained, a huge time scale that is prohibitive when one needs to destabilize emulsions in industrial applications, e.g. in oil recovery. We therefore investigate the squeezing technique as an alternative method for destabilization. We will show below that in spite of the large stability of our emulsion, the squeezing technique allows to readily and completely destabilize our emulsion on a much shorter time scale. We apply this to the most difficult case: very concentrated and very stable emulsions, and successfully destabilize these.

\subsection{Squeeze flow}
\begin{figure}
	\centering
	\includegraphics[width=\linewidth]{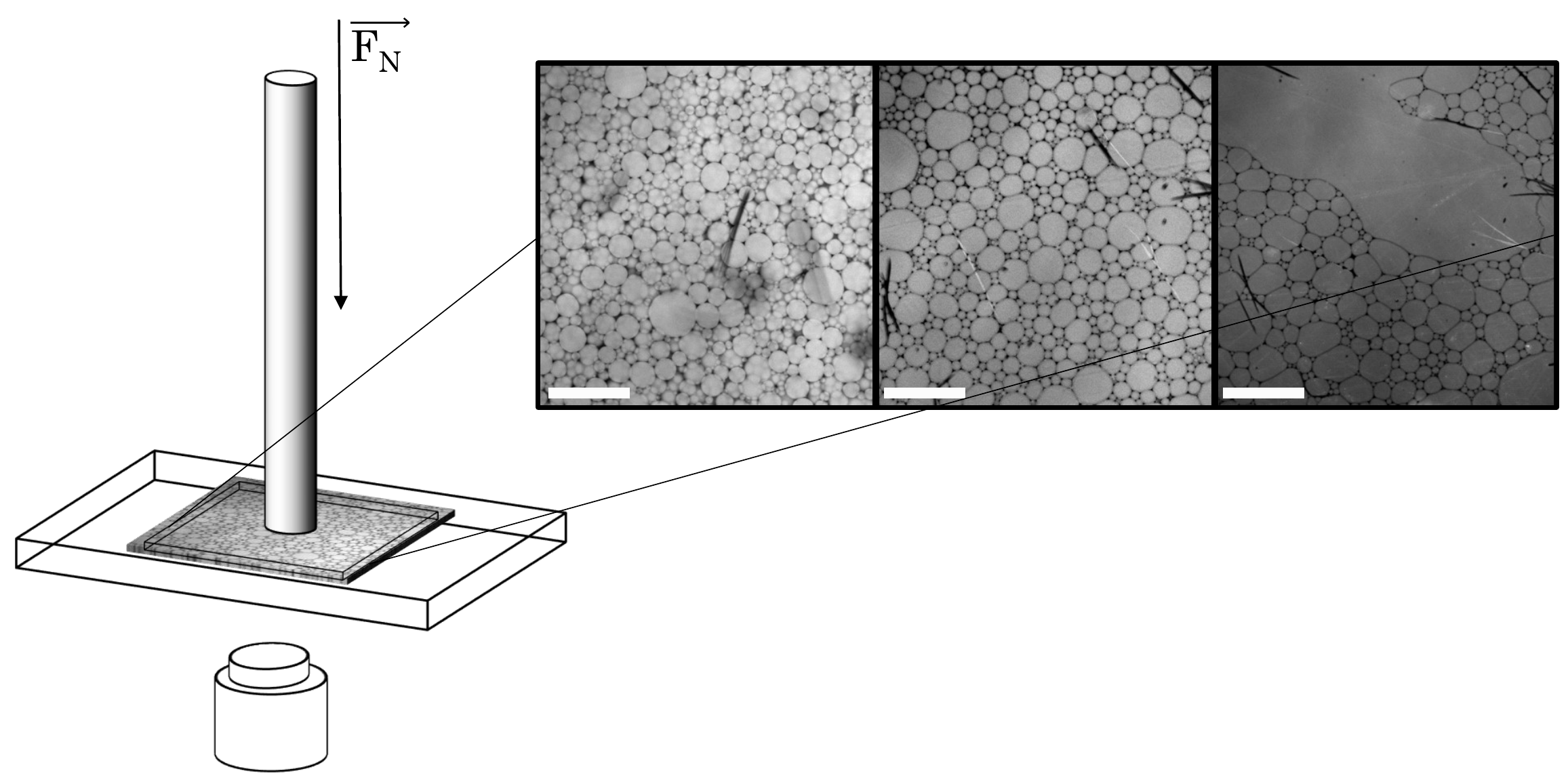}
	\caption{Setup (left, not to scale) and typical microscopy images (right, dispersed phase rendered bright) of emulsion destabilization by squeezing. Typical confocal images reveal, from left to right: strong deformations of the emulsion occur until a critical thickness is reached, after which coalescence events in the emulsion are observed. Scale bars correspond to 50 $\mu$m.} 
	\label{squeeze_setup}
\end{figure}

To destabilize our highly-concentrated silicone oil-in-glycerol/water emulsion stabilized by sodium dodecyl sulfate, we squeeze it between two glass plates by lowering the position of a rheometer head that is placed on top of a confocal microscope, as illustrated in Figure \ref{squeeze_setup}. Also shown in the Figure are three stages of destabilization. First, the emulsion layer is still larger than the average drop diameter and the droplets are spherical. The oil droplets have an average droplet diameter of 20 $\mu$m with a dispersity of 20 \%. This rather high dispersity is the result of using a standard emulsifier. The emulsions studied here are representative for emulsions encountered in for example the oil recovery process and in food production. Then, the sample thickness is decreased, resulting in one layer of emulsion droplets that deform to polygonal shapes with very thin layers of water in between the droplets. When the emulsion is squeezed further, coalescence events start to occur and the emulsion becomes unstable.

\begin{figure*}
	\centering
	\includegraphics[width=\linewidth]{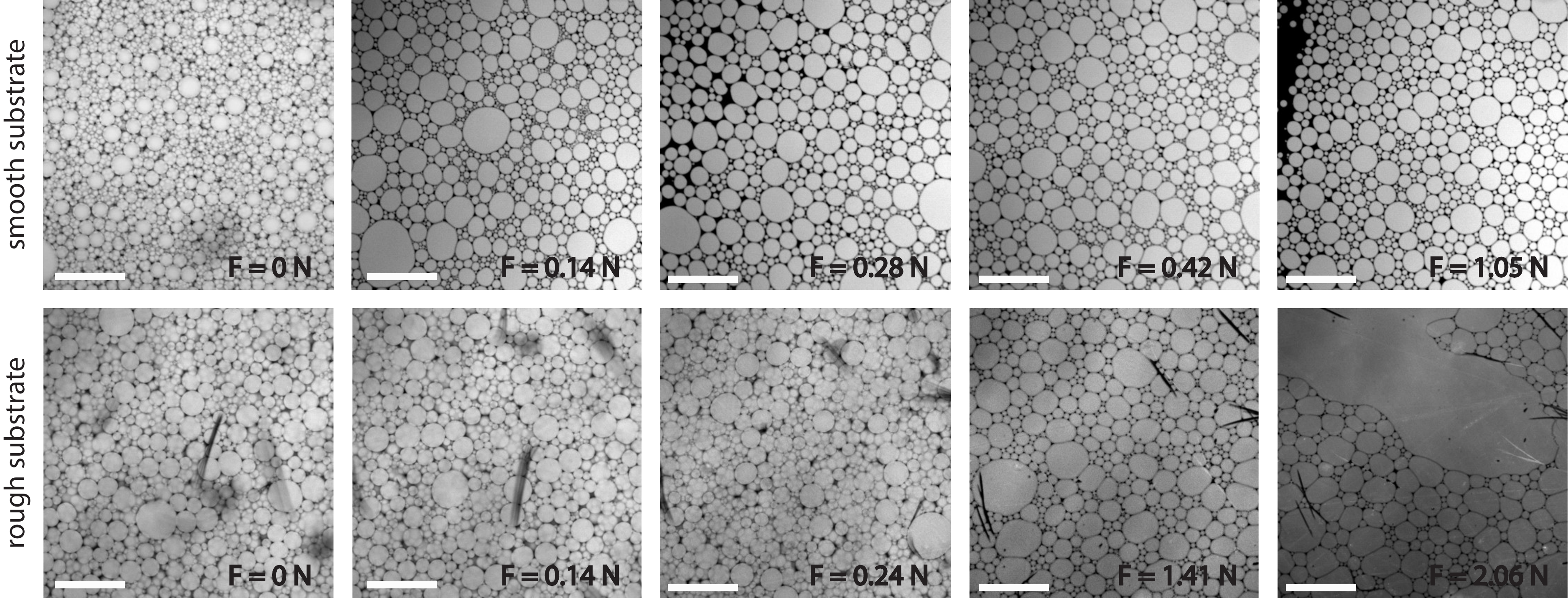}
	\caption{Squeezing experiments on a smooth (top) and a rough (bottom) glass substrate. From left to right, the squeezing force increases from 0 to 1.05 N. Whereas the oil droplets glide over the smooth substrate, they are pinned on the rough substrate, allowing for further deformation of the droplets and finally destabilization of the emulsion. Scale bars are 50 $\mu$m.} 
	\label{roughsmoothsubstrate}
\end{figure*}

The properties of the glass plate onto which the emulsion is squeezed are of paramount importance for the destabilization behavior. In Figure \ref{roughsmoothsubstrate}, we compare a smooth glass substrate (top row) with a rough glass substrate (bottom row). The rough glass substrate has a grit designation of P800, which refers to a roughness of 22 $\mu$m. We choose this specific grit designation, as the roughness of the glass substrate is very close to the average droplet diameter in our emulsions. 

Only slightly compressing the emulsion does not reveal differences. However, when the emulsion is squeezed further, the smooth surface allows oil droplets to slide over the substrate, eventually being pushed out from between the glass plates, without inducing coalescence of oil droplets, while the rough surface  prevents gliding of the oil droplets, causing only the water to be squeezed out from the emulsion. This results in very thin films between the oil droplets that can break and thereby induce coalescence events. An elaborate investigation in the reason behind these coalescence events that result in emulsion destabilization can be found in \cite{Dekker2020}.

We use this rough plate in experiments with emulsions having oil droplets between 10 and 50 $\mu$m \cite{Dekker2020}, where the rough plate successfully immobilizes the oil droplets. This shows that the grit designation does not have to perfectly match the average droplet diameter to immobilize the emulsion droplets. However, when much smaller or much larger oil droplets need to be immobilized, a different grit designation will most probably work better. A more elaborate study on the importance of surface roughness and surface chemistry on immobilizing oil droplets and preventing wall slip can be found in \cite{Paredes2015}. Furthermore, follow-up research on this topic focuses on the morphology and wettability of the surfaces used in the oil industry. Our results give a good indication that finding the right morphology and wettability can greatly help in destabilizing the emulsions during oil recovery.

\begin{figure}
	\centering
	\includegraphics[width=\linewidth]{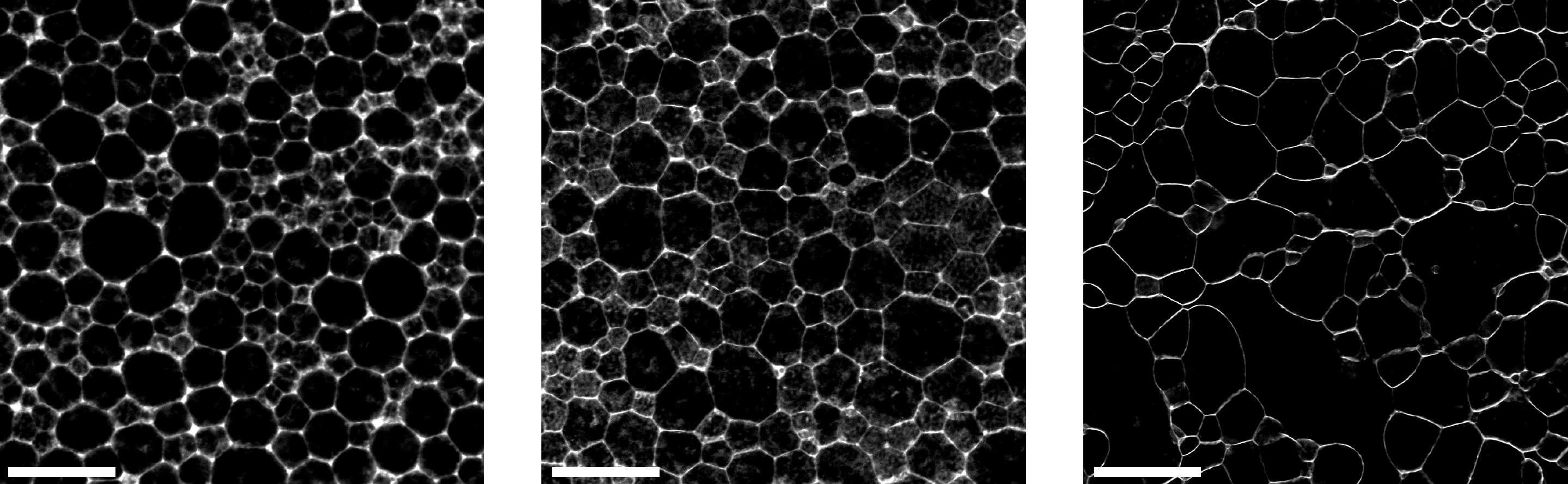}
	\caption{BSA/PGA-stabilized emulsion destabilization by squeezing. Confocal images showing deformation of the emulsion droplets and coalescence events in the emulsion after a critical sample thickness is reached. Scale bars are 100 $\mu$m.} 
	\label{BSAemulsion}
\end{figure}

To validate this new emulsion destabilization technique, various emulsions have been tested. Silicone oil-in-water emulsions with different oil droplet sizes, ranging from 10 to 50 $\mu$m, have been discussed in \cite{Dekker2020}. All emulsions successfully destabilize in the same way as presented here. Variations in the stabilizing agent have been investigated as well. In Figure \ref{BSAemulsion}, we show the results of a silicone oil-in-water emulsion stabilized by BSA and PGA. The destabilization behavior is very similar to the emulsion shown in Figure \ref{squeeze_setup}. The confocal images show the deformation of the oil droplets into polygonal shapes, the increase of the oil/water ratio and finally film rupture leading to coalescence events of two droplets. However, the whole process is somewhat slower, due to the formation of a rigid interface in the presence of BSA and PGA. This rigid interface increases the droplet's resistance to deformation and slows down drainage of water. 

\subsection{Measurements of the squeezing force}
\begin{figure}
	\centering
	\includegraphics[width=\linewidth]{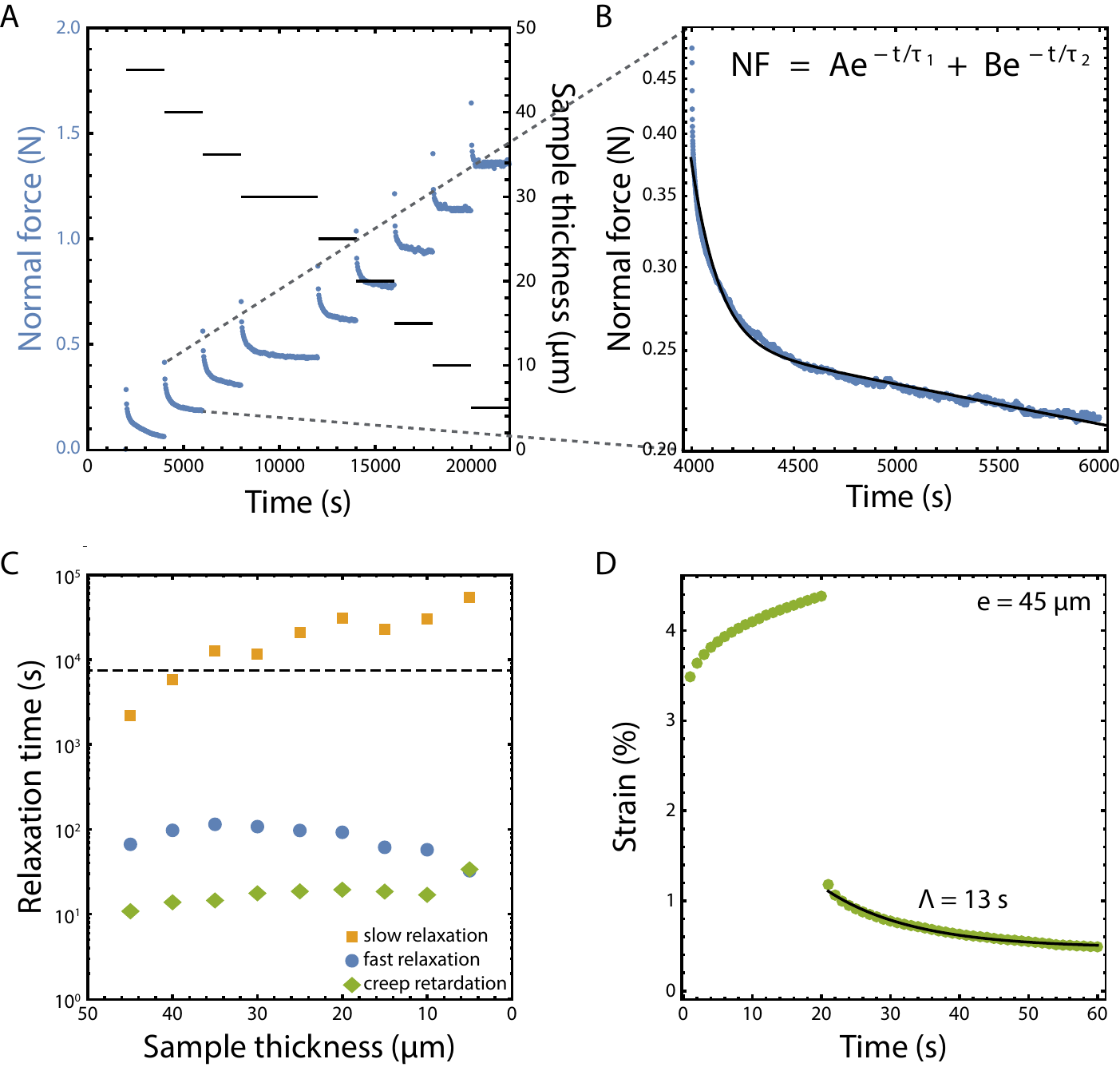}
	\caption{Squeezing force measurements. (A) Normal force and film thickness as a function of time during step wise increments of the squeezing force. (B) Normal force relaxation fitted by a bi-exponential (solid line). (C) Relaxation times and creep retardation times as a function of the sample thickness. The dashed line is the theoretical prediction for the characteristic time of water flow dynamics. (D) Creep experiment for a sample thickness of 45 $\mu$m. }
	\label{relaxationtime}
\end{figure}

We now look at the time-dependence of the forces that are necessary to squeeze and finally destabilize the emulsion. To gain insight into the dynamics of droplet formation and coalescence, we measure the squeezing force during destabilization experiments. Figure \ref{relaxationtime}A shows the results of a typical experiment. Each time the gap size is reduced, the normal force increases, followed by a time-dependent relaxation: a fast relaxation in approximately the first 500 s, followed by a slow relaxation. We perform a bi-exponential fitting of the normal force for each sample thickness, see Figure \ref{relaxationtime}B. From this bi-exponential fitting, two relaxation times can be obtained: $\tau_1$ on the order of 100 s, and $\tau_2$ on the order of $10^4$ s. The relaxation times are plotted versus the sample thickness in Figure \ref{relaxationtime}C. 

It is worthwhile noting that that performing rheology and squeezing experiments on confined systems such as the one discussed here is not always trivial. The alignment of the rheometer has to be very good in order to obtain reproducible data, since any misalignment can generate secondary flows. For the rheometer that we use, the most sensitive test of the quality of the alignment is doing standard rheology with a very small gap. The truncation of one of our 50 mm cone-plate geometries is 15 $\mu$m, and this can be positioned with sufficient accuracy to obtain very reproducible results. Rheology on high volume fraction oil-in-water emulsions \cite{Paredes2013} using this gemoetry showed the formation of a master curve when scaling the shear rate and shear stress with respect to the distance to jamming. This allows predicting the flow properties of our highly concentrated emulsions from the average droplet size and the volume fraction of dispersed phase but also shows the high degree of reproducibility. 

\subsection{Creep and Darcy flow timescales}
The slower of the two relaxation processes, with relaxation times around $10^4$ s, can be ascribed to the characteristic time for the water to squeeze out of the emulsion, similar to what is observed for very open bio-polymer networks\cite{DeCagny2016}. Here, elastic stresses drive the liquid out, and the fluid resistance is of the Darcy type, leading to

\begin{equation}
\tau \simeq \frac{\eta R_0^2}{\phi^3 G_0 \xi^2},
\label{darcy}
\end{equation}

with $\eta$ the viscosity of the continuous phase, $R_0$ the size of the microscopy glass plate, $\phi$ the volume fraction of dispersed phase, $G_0$ the shear modulus and $\xi$ the pore size between the oil droplets \cite{DeCagny2016}, with $\eta = 1$ mPa$\cdot$s, $R_0 = 20$ mm, and $G_0\approx G' = 145$ Pa (measured on an Anton Paar MCR 302 rheometer using a CP50 geometry at a strain of 1\%). The volume fraction of the dispersed phase and the pore size between the oil droplets strongly depend on the sample thickness. As discussed in \cite{Dekker2020}, the volume fraction increases from 0.80 to almost 0.95 when coalescence events start to occur. Due to the water being squeezed out of the emulsion, the pore size between the oil droplets is much smaller at a sample thickness of 5 $\mu$m than at a sample thickness of 50 $\mu$m. From the confocal images we estimate that $\xi \approx 2$ $\mu$m at the start of the experiment and decreases roughly by a factor of 10. We therefore obtain a relaxation time $\tau \approx 1.5 \cdot 10^3$ s at $e = 50$ $\mu$m and $\tau \approx 8.0 \cdot 10^4$ s at $e = 5$ $\mu$m. The relaxation time is indicated with a dashed line in Figure \ref{relaxationtime}C. We assume that the relaxation time increases linearly with decreasing sample thickness. The predicted Darcy flow times are in reasonable agreement with the slow relaxation times from the squeeze experiments. 

To elucidate the fast relaxation process during squeezing, creep measurements in simple shear are performed. These measurements can give insight in the viscoplastic properties of the material. The emulsion is deformed for a period of 20 s with a shear stress of 5 Pa, and then allowed to recover over a period of 40 s. This results in a creep and creep recovery curve, as can be seen in Figure \ref{relaxationtime}D. The behaviour during creep and creep recovery can be analysed using Burgers' model \cite{Burgers1935,Mezger2011}. This model is a combination of the Maxwell model and the Kelvin-Voigt model \cite{Schmalholz2001}. With this model, the creep recovery curve can be fitted to obtain a characteristic retardation time

\begin{equation}
\gamma (t) = \gamma_{\textup{max}}-\left(\frac{\tau_0}{G_1}\right)-\left(\frac{\tau_0}{G_2}\right)[1-\exp{\left(\frac{-t}{\Lambda}\right)}],
\end{equation}

where $G_1$ and $G_2$ are the elastic moduli of both springs in the Burgers' model, $\tau_0$ is the applied stress during creep (in our case $\tau_0$ = 5 Pa), and $\Lambda$ is the retardation time of creep recovery. Figure \ref{relaxationtime}D shows the fit of the creep recovery curve according to the Burgers' model, yielding a retardation time of 13 s for a layer thickness of 45 $\mu$m. The creep time for each layer thickness is shown in Figure \ref{relaxationtime}C (green diamonds). We find that the faster relaxation time in squeeze flow and the retardation time obtained from creep measurements are similar. This indicates that the faster relaxation time observed in squeeze flow measurements is due to the recovery of the system after deformation. Differences between the times are likely due to the fact that, in squeeze flow, the emulsion is deformed in the direction perpendicular to the plates, whereas in creep the emulsion is deformed in the direction parallel to the plates. However, in both measurements the creep of the emulsion resulting from a deformation is probed. 

Compared to the drainage of water on a time scale of $10^6$ s, this process can be sped up significantly by compressing the oil droplets between two glass plates. Furthermore, creep on a time scale of roughly 100 s is also observed, a process that does not take place in a stationary emulsion like in Figure \ref{hartland}. Creep deformation of the emulsion leads to rearrangements of the oil droplets. Water drainage causes thinning of the films between the droplets. Together, these phenomena lead to coalescence events and finally emulsion destabilization. 

\subsection{Squeezing force due to the Laplace pressure in the drops}
In order to quantitatively understand the normal force we measured during the squeezing experiments, we have developed a model that considers the squeezing of individual droplets.

A stable, spherical droplet of diameter $D$ has a pressure gradient over its interface given by the Laplace pressure, resulting from the curvature of its surface:

\begin{equation}
    \Delta P = \frac{4\gamma}{D}
\end{equation}

with $\gamma$ the interfacial tension of the system. We assume that the force required to squeeze a droplet, is directly related to this pressure, and that this force is roughly equal to $\Delta P$ times the cross-sectional area of the droplet $\frac{\pi D^2}{4}$. Furthermore, it is implied that only the droplets with a diameter larger than sample thickness $e$ are actually squeezed and yield a normal force:

\[ f_{droplet}(D) =
   \begin{cases}
     \pi \gamma D & \text{if $D \geq e$} \\
     0 & \text{if $D < e$}
   \end{cases}
\]

\begin{figure}
	\centering
	\includegraphics[width=\linewidth]{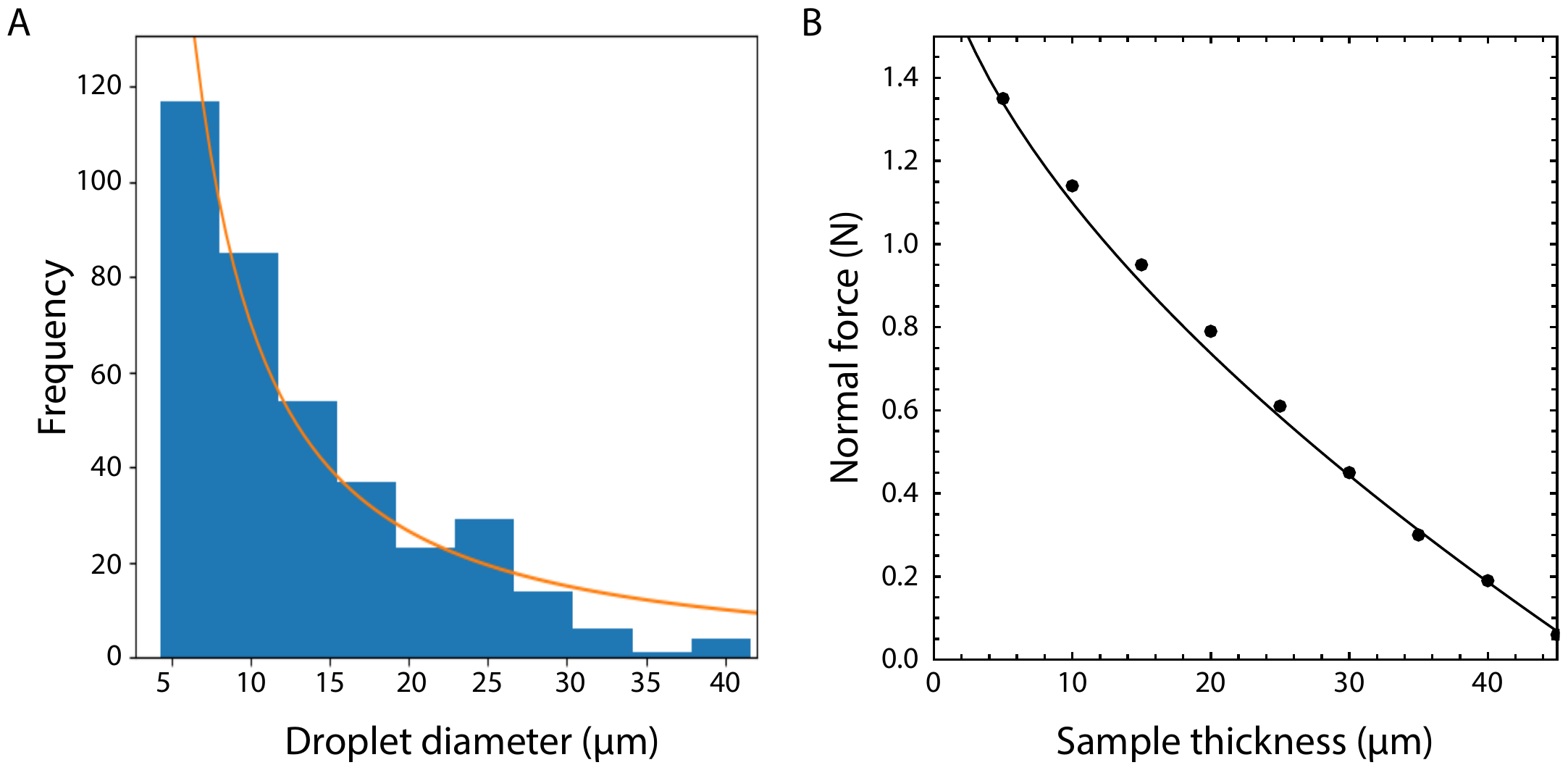}
	\caption{Relating the Laplace pressure to the normal forces during squeezing. (A) Distribution of droplet sizes in an emulsion before squeezing. (B) Experimental normal forces during a squeeze experiment (circles) and a fit based on the droplet size distribution and Laplace pressures of these droplets (solid line).}
	\label{laplaceforce}
\end{figure}

To find the total normal force exerted by all droplets in the emulsion, it is required to know the size distribution and the number of droplets in the total area squeezed. This size distribution is obtained from a confocal image before squeezing. Droplets with diameter between 5 of 40 $\mu$m are found, according to the distribution plotted in Figure \ref{laplaceforce}A. It can be noted that a high number of small droplets (between 5 and 15 $\mu$m) is found. This is partly a result of the confocal image showing only a single plane, whereas not all droplets are positioned at the same height. In order to maintain the simplicity of the model, we decided not to compensate for this optical effect. The solid line in Figure \ref{laplaceforce}A shows a power-law fit of the droplet size distribution. Considering that the distribution found in the confocal images is uniform underneath the entire squeezing area of 20 x 20 mm$^2$, we estimate that droplet distribution $v(D)$ is given by:

\begin{equation}
   v(D) = \beta \cdot \left( \frac{D}{D_0}\right)^{-x} 
\label{distribution}
\end{equation}

with $\beta = 1.67 \cdot 10^5$ $\mu$m$^{-1}$, $D_0 = 10$ $\mu$m and $x = 1.5$.The estimated total number of droplets in the squeezing area with diameter between $D$ and $D + dD$ is thus approximated by $v(D) \cdot dD$. 

The total force to squeeze the emulsion, $F_{tot}$, is the sum of the distribution of droplets times the Laplace force of a single droplet:

\begin{equation}
\begin{split}
    F_{tot} = \int_{D=e}^{D=D_{m}}\!f_{droplet}(D)\cdot v(D) \cdot dD \\
    = \pi \beta \gamma \int_{D=e}^{D=D_{m}}\! D \left(\frac{D}{D_0}\right)^{-x} dD 
\end{split}
\end{equation}

Notice that diameters smaller than the sample thickness, $D < e$ do not have to be accounted for in the integral, as these droplets do not contribute to the force. The upper limit of the integral is manually limited by a maximum diameter $D_m$ meaning that the no droplets with diameter larger than $D_m$ are present in the emulsion. Evaluation of the integral yields the result:

\begin{equation}
    F_{tot} = 
    \frac{\pi}{2-x} \beta \gamma D_0^2 \left(\left(\frac{D_m}{D_0}\right)^{2-x}-\left(\frac{e}{D_0}\right)^{2-x}\right)
\label{totalforce}
\end{equation}

Figure \ref{laplaceforce}B shows that there is good agreement between this formula and the measurements. The circles show the experimental normal force as a function of the sample thickness, retrieved from the data in Figure \ref{relaxationtime}A. The solid line shows equation \ref{totalforce}, with the parameters $D_m = 48$ $\mu$m, and $\gamma = 9$ mN/m. To validate this interfacial tension $\gamma$, we perform pendant drop measurements on our sample. These measurements yield an interfacial tension of 10 $\pm$ 1.5 mN/m, within the experimental accuracy of the value obtained from the model fit. 

We are aware that this model is still very simplified and that certain parameters should ideally be determined more precisely, like the exact shape of the distribution, and the maximum droplet diameter. However, despite its simplicity, the theory does link the Laplace pressure within single droplets to the normal forces encountered during a squeezing experiment in a way that predicts the correct trend and the correct order of magnitude. This model is a first step to estimate the squeeze force necessary to destabilize the emulsion from the distribution of droplet sizes. Further research is necessary in order to see whether the model still holds for a system with completely different droplet sizes. This is beyond the scope of the current paper. 

\subsection{Scaling up the destabilization technique}
Our squeeze flow technique to destabilize emulsions gives opportunities in the oil industry for crude oil recovery in a surfactant-free way. However, we are aware of the various complexities that arise when going from a lab environment with only a few milliliters of sample to the field where tons of liters have to be processed. This project is therefore performed in close collaboration with Shell, where more research will be performed to scale up our technique to destabilize emulsions. Despite the flow effects that start to play a role on a large scale, we believe that the pressures and timescales that we measure are comparable to the values in the field. Therefore, these results are a big step in the right direction to use this new, unique technique on a large scale.

Besides, the technique is very suitable for fast screening of emulsions in very small quantities. This for example allows for screening of the effect of emulsifiers on the coalescence properties. The great advantages of testing on small volumes, e.g. reduction in costs and time and less safety issues, are the reason for companies to have interest in testing destabilization techniques on small volumes. Understanding what happens on a microscopic scale is of great importance for crude oil recovery \cite{Chatzigiannakis2020}.

\section{Conclusion}
In summary, we studied the time scales for emulsion destabilization in highly-concentrated oil-in-water emulsions induced by the squeeze flow technique. Whereas normally water drainage only occurs in oil-in-water emulsions with oil fractions below the jamming point, this process can be induced and sped up by squeezing the emulsion between two glass plates. Coalescence events eventually occur at high oil volume fractions with very thin layers of continuous phase between the droplets. The emulsion used in this paper differs slightly from the emulsions used by Dekker \textit{et al.}\cite{Dekker2020} In the latter paper, the continuous phases contained only water with 1 wt\% of SDS and Rhodamine B was added as a dye to the continuous phase. Despite these small differences, the observed behavior in squeeze experiments is the same, indicating that the results reported here are quite general. We show that the coalescence process is strongly affected by the roughness of the glass substrate, with rough surfaces leading to pinning and coalescence of the oil droplets, while smooth surfaces allow the emulsion to glide without coalescence. 

Force relaxation measurements highlight two important processes during squeezing of an emulsion. First, drainage of water on a time scale of $10^4$ s, which leads to thinning of the films between the oil droplets. Second, creep deformation on a time scale of $10^2$ s, which leads to rearrangements processes in the oil droplets. These two processes together result in coalescence events and then full destabilization of the emulsion. We propose a model for the total force that needs to be exerted on the emulsion to squeeze the emulsion that is based on the Laplace pressure of single oil droplets and the size distribution of these droplets. This model allows us to predict the necessary force to induce destabilization from the droplet size distribution in an emulsion. The acquired insights in the time scales for creep deformation and water drainage and the squeeze forces can be used to guide the scaling up of the process towards use in large-scale crude oil recovery.

\begin{acknowledgments}
This work is part of the research program Controlling Multiphase Flow with project number 680-91-012, which is (partly) financed by the Dutch Research Council (NWO) and co-funded by TKI-E\&I with the supplementary grant 'TKI- Toeslag' for Topconsortia for Knowledge and Innovation (TKI’s) of the Ministry of Economic Affairs and Climate Policy. This work took place within the framework of the Institute of Sustainable Process Technology. The authors thank the workshop of the University of Amsterdam for their technical assistance. A.D acknowledges funding from the European Union’s Horizon 2020 research and innovation program under the Individual Marie Skłodowska-Curie fellowship grant agreement number 798455. This work was partially funded by Evodos, Shell Global Solutions International B.V. and Unilever R\&D.
\end{acknowledgments}

\section*{data availability}
The data that support the findings of this study are available from the corresponding author upon reasonable request.


\begin{thebibliography}{41}%
\makeatletter
\providecommand \@ifxundefined [1]{%
 \@ifx{#1\undefined}
}%
\providecommand \@ifnum [1]{%
 \ifnum #1\expandafter \@firstoftwo
 \else \expandafter \@secondoftwo
 \fi
}%
\providecommand \@ifx [1]{%
 \ifx #1\expandafter \@firstoftwo
 \else \expandafter \@secondoftwo
 \fi
}%
\providecommand \natexlab [1]{#1}%
\providecommand \enquote  [1]{``#1''}%
\providecommand \bibnamefont  [1]{#1}%
\providecommand \bibfnamefont [1]{#1}%
\providecommand \citenamefont [1]{#1}%
\providecommand \href@noop [0]{\@secondoftwo}%
\providecommand \href [0]{\begingroup \@sanitize@url \@href}%
\providecommand \@href[1]{\@@startlink{#1}\@@href}%
\providecommand \@@href[1]{\endgroup#1\@@endlink}%
\providecommand \@sanitize@url [0]{\catcode `\\12\catcode `\$12\catcode
  `\&12\catcode `\#12\catcode `\^12\catcode `\_12\catcode `\%12\relax}%
\providecommand \@@startlink[1]{}%
\providecommand \@@endlink[0]{}%
\providecommand \url  [0]{\begingroup\@sanitize@url \@url }%
\providecommand \@url [1]{\endgroup\@href {#1}{\urlprefix }}%
\providecommand \urlprefix  [0]{URL }%
\providecommand \Eprint [0]{\href }%
\providecommand \doibase [0]{https://doi.org/}%
\providecommand \selectlanguage [0]{\@gobble}%
\providecommand \bibinfo  [0]{\@secondoftwo}%
\providecommand \bibfield  [0]{\@secondoftwo}%
\providecommand \translation [1]{[#1]}%
\providecommand \BibitemOpen [0]{}%
\providecommand \bibitemStop [0]{}%
\providecommand \bibitemNoStop [0]{.\EOS\space}%
\providecommand \EOS [0]{\spacefactor3000\relax}%
\providecommand \BibitemShut  [1]{\csname bibitem#1\endcsname}%
\let\auto@bib@innerbib\@empty
\bibitem [{\citenamefont {Leal-Calderon}, \citenamefont {Schmitt},\ and\
  \citenamefont {Bibette}(2007)}]{Leal-Calderon2007}%
  \BibitemOpen
  \bibfield  {author} {\bibinfo {author} {\bibfnamefont {F.}~\bibnamefont
  {Leal-Calderon}}, \bibinfo {author} {\bibfnamefont {V.}~\bibnamefont
  {Schmitt}},\ and\ \bibinfo {author} {\bibfnamefont {J.}~\bibnamefont
  {Bibette}},\ }\href@noop {} {\emph {\bibinfo {title} {{Emulsion Science basic
  principles}}}}\ (\bibinfo  {publisher} {Springer},\ \bibinfo {year} {2007})\
  p.\ \bibinfo {pages} {225}\BibitemShut {NoStop}%
\bibitem [{\citenamefont {Bibette}\ \emph {et~al.}(1992)\citenamefont
  {Bibette}, \citenamefont {Morse}, \citenamefont {Witten},\ and\ \citenamefont
  {Weitz}}]{Bibette1992}%
  \BibitemOpen
  \bibfield  {author} {\bibinfo {author} {\bibfnamefont {J.}~\bibnamefont
  {Bibette}}, \bibinfo {author} {\bibfnamefont {D.~C.}\ \bibnamefont {Morse}},
  \bibinfo {author} {\bibfnamefont {T.~A.}\ \bibnamefont {Witten}},\ and\
  \bibinfo {author} {\bibfnamefont {D.~A.}\ \bibnamefont {Weitz}},\ }\bibfield
  {title} {\enquote {\bibinfo {title} {{Stability Criteria for Emulsions}},}\
  }\href@noop {} {\bibfield  {journal} {\bibinfo  {journal} {Physical Review
  Letters}\ }\textbf {\bibinfo {volume} {69}},\ \bibinfo {pages} {2439--2443}
  (\bibinfo {year} {1992})}\BibitemShut {NoStop}%
\bibitem [{\citenamefont {Wasan}, \citenamefont {Nikolov},\ and\ \citenamefont
  {Aimetti}(2004)}]{Wasan2004}%
  \BibitemOpen
  \bibfield  {author} {\bibinfo {author} {\bibfnamefont {D.~T.}\ \bibnamefont
  {Wasan}}, \bibinfo {author} {\bibfnamefont {A.~D.}\ \bibnamefont {Nikolov}},\
  and\ \bibinfo {author} {\bibfnamefont {F.}~\bibnamefont {Aimetti}},\
  }\bibfield  {title} {\enquote {\bibinfo {title} {{Nanostructured hybrid
  materials from aqueous polymer dispersions}},}\ }\href
  {https://doi.org/10.1016/j.cis.2003.10.017} {\bibfield  {journal} {\bibinfo
  {journal} {Advances in Colloid and Interface Science}\ }\textbf {\bibinfo
  {volume} {108-109}},\ \bibinfo {pages} {187--195} (\bibinfo {year}
  {2004})}\BibitemShut {NoStop}%
\bibitem [{\citenamefont {Bergenst{\aa}hl}\ and\ \citenamefont
  {Robins}(1998)}]{Bergenstahl1998}%
  \BibitemOpen
  \bibfield  {author} {\bibinfo {author} {\bibfnamefont {B.}~\bibnamefont
  {Bergenst{\aa}hl}}\ and\ \bibinfo {author} {\bibfnamefont {M.}~\bibnamefont
  {Robins}},\ }\bibfield  {title} {\enquote {\bibinfo {title} {{Food colloids,
  emulsions, gels and foams -- Desirable entanglements - food polymers and
  product stability}},}\ }\href {https://doi.org/10.1016/S1359-0294(98)80090-3}
  {\bibfield  {journal} {\bibinfo  {journal} {Current Opinion in Colloid and
  Interface Science}\ }\textbf {\bibinfo {volume} {3}},\ \bibinfo {pages}
  {625--626} (\bibinfo {year} {1998})}\BibitemShut {NoStop}%
\bibitem [{\citenamefont {Glampedaki}\ and\ \citenamefont
  {Dutschk}(2014)}]{Glampedaki2014}%
  \BibitemOpen
  \bibfield  {author} {\bibinfo {author} {\bibfnamefont {P.}~\bibnamefont
  {Glampedaki}}\ and\ \bibinfo {author} {\bibfnamefont {V.}~\bibnamefont
  {Dutschk}},\ }\bibfield  {title} {\enquote {\bibinfo {title} {{Stability
  studies of cosmetic emulsions prepared from natural products such as wine,
  grape seed oil and mastic resin}},}\ }\href
  {https://doi.org/10.1016/j.colsurfa.2014.02.048} {\bibfield  {journal}
  {\bibinfo  {journal} {Colloids and Surfaces A: Physicochemical and
  Engineering Aspects}\ }\textbf {\bibinfo {volume} {460}},\ \bibinfo {pages}
  {306--311} (\bibinfo {year} {2014})}\BibitemShut {NoStop}%
\bibitem [{\citenamefont {Degrand}, \citenamefont {Michon},\ and\ \citenamefont
  {Bosc}(2016)}]{Degrand2016}%
  \BibitemOpen
  \bibfield  {author} {\bibinfo {author} {\bibfnamefont {L.}~\bibnamefont
  {Degrand}}, \bibinfo {author} {\bibfnamefont {C.}~\bibnamefont {Michon}},\
  and\ \bibinfo {author} {\bibfnamefont {V.}~\bibnamefont {Bosc}},\ }\bibfield
  {title} {\enquote {\bibinfo {title} {{New insights into the study of the
  destabilization of oil-in-water emulsions with dextran sulfate provided by
  the use of light scattering methods}},}\ }\href
  {https://doi.org/10.1016/j.foodhyd.2015.08.021} {\bibfield  {journal}
  {\bibinfo  {journal} {Food Hydrocolloids}\ }\textbf {\bibinfo {volume}
  {52}},\ \bibinfo {pages} {848--856} (\bibinfo {year} {2016})}\BibitemShut
  {NoStop}%
\bibitem [{\citenamefont {White}\ and\ \citenamefont
  {Nitsch}(2000)}]{White2000}%
  \BibitemOpen
  \bibfield  {author} {\bibinfo {author} {\bibfnamefont {L.~S.}\ \bibnamefont
  {White}}\ and\ \bibinfo {author} {\bibfnamefont {A.~R.}\ \bibnamefont
  {Nitsch}},\ }\bibfield  {title} {\enquote {\bibinfo {title} {{Solvent
  recovery from lube oil filtrates with a polyimide membrane}},}\ }\href@noop
  {} {\bibfield  {journal} {\bibinfo  {journal} {Journal of Membrane Science}\
  }\textbf {\bibinfo {volume} {179}},\ \bibinfo {pages} {267--274} (\bibinfo
  {year} {2000})}\BibitemShut {NoStop}%
\bibitem [{\citenamefont {Pe{\~{n}}a}, \citenamefont {Hirasaki},\ and\
  \citenamefont {Miller}(2005)}]{Pena2005}%
  \BibitemOpen
  \bibfield  {author} {\bibinfo {author} {\bibfnamefont {A.~A.}\ \bibnamefont
  {Pe{\~{n}}a}}, \bibinfo {author} {\bibfnamefont {G.~J.}\ \bibnamefont
  {Hirasaki}},\ and\ \bibinfo {author} {\bibfnamefont {C.~A.}\ \bibnamefont
  {Miller}},\ }\bibfield  {title} {\enquote {\bibinfo {title} {{Chemically
  Induced Destabilization of Water-in-Crude Oil Emulsions}},}\ }\href
  {https://doi.org/10.1021/ie049666i} {\bibfield  {journal} {\bibinfo
  {journal} {Industrial and Engineering Chemistry Research}\ }\textbf {\bibinfo
  {volume} {44}},\ \bibinfo {pages} {1139--1149} (\bibinfo {year}
  {2005})}\BibitemShut {NoStop}%
\bibitem [{\citenamefont {Less}\ \emph {et~al.}(2008)\citenamefont {Less},
  \citenamefont {Hannisdal}, \citenamefont {Bj{\o}rklund},\ and\ \citenamefont
  {Sj{\"{o}}blom}}]{Less2008}%
  \BibitemOpen
  \bibfield  {author} {\bibinfo {author} {\bibfnamefont {S.}~\bibnamefont
  {Less}}, \bibinfo {author} {\bibfnamefont {A.}~\bibnamefont {Hannisdal}},
  \bibinfo {author} {\bibfnamefont {E.}~\bibnamefont {Bj{\o}rklund}},\ and\
  \bibinfo {author} {\bibfnamefont {J.}~\bibnamefont {Sj{\"{o}}blom}},\
  }\bibfield  {title} {\enquote {\bibinfo {title} {{Electrostatic
  destabilization of water-in-crude oil emulsions : Application to a real case
  and evaluation of the Aibel VIEC technology}},}\ }\href
  {https://doi.org/10.1016/j.fuel.2008.03.004} {\bibfield  {journal} {\bibinfo
  {journal} {Fuel}\ }\textbf {\bibinfo {volume} {87}},\ \bibinfo {pages}
  {2572--2581} (\bibinfo {year} {2008})}\BibitemShut {NoStop}%
\bibitem [{\citenamefont {Katepalli}\ \emph {et~al.}(2016)\citenamefont
  {Katepalli}, \citenamefont {Bose}, \citenamefont {Hatton},\ and\
  \citenamefont {Blankschtein}}]{Katepalli2016}%
  \BibitemOpen
  \bibfield  {author} {\bibinfo {author} {\bibfnamefont {H.}~\bibnamefont
  {Katepalli}}, \bibinfo {author} {\bibfnamefont {A.}~\bibnamefont {Bose}},
  \bibinfo {author} {\bibfnamefont {T.~A.}\ \bibnamefont {Hatton}},\ and\
  \bibinfo {author} {\bibfnamefont {D.}~\bibnamefont {Blankschtein}},\
  }\bibfield  {title} {\enquote {\bibinfo {title} {{Destabilization of
  Oil-in-Water Emulsions Stabilized by Non-ionic Surfactants: Effect of
  Particle Hydrophilicity}},}\ }\href
  {https://doi.org/10.1021/acs.langmuir.6b03289} {\bibfield  {journal}
  {\bibinfo  {journal} {Langmuir}\ }\textbf {\bibinfo {volume} {32}},\ \bibinfo
  {pages} {10694--10698} (\bibinfo {year} {2016})}\BibitemShut {NoStop}%
\bibitem [{\citenamefont {Djuve}\ \emph {et~al.}(2001)\citenamefont {Djuve},
  \citenamefont {Yang}, \citenamefont {Fjellanger}, \citenamefont
  {Sj{\"{o}}blom},\ and\ \citenamefont {Pelizzetti}}]{Djuve2001}%
  \BibitemOpen
  \bibfield  {author} {\bibinfo {author} {\bibfnamefont {J.}~\bibnamefont
  {Djuve}}, \bibinfo {author} {\bibfnamefont {X.}~\bibnamefont {Yang}},
  \bibinfo {author} {\bibfnamefont {I.~J.}\ \bibnamefont {Fjellanger}},
  \bibinfo {author} {\bibfnamefont {J.}~\bibnamefont {Sj{\"{o}}blom}},\ and\
  \bibinfo {author} {\bibfnamefont {E.}~\bibnamefont {Pelizzetti}},\ }\bibfield
   {title} {\enquote {\bibinfo {title} {{Chemical destabilization of crude oil
  based emulsions and asphaltene stabilized emulsions}},}\ }\href
  {https://doi.org/10.1007/s003960000413} {\bibfield  {journal} {\bibinfo
  {journal} {Colloid and Polymer Science}\ }\textbf {\bibinfo {volume} {279}},\
  \bibinfo {pages} {232--239} (\bibinfo {year} {2001})}\BibitemShut {NoStop}%
\bibitem [{\citenamefont {Fan}, \citenamefont {Simon},\ and\ \citenamefont
  {Sj{\"{o}}blom}(2009)}]{Fan2009}%
  \BibitemOpen
  \bibfield  {author} {\bibinfo {author} {\bibfnamefont {Y.}~\bibnamefont
  {Fan}}, \bibinfo {author} {\bibfnamefont {S.}~\bibnamefont {Simon}},\ and\
  \bibinfo {author} {\bibfnamefont {J.}~\bibnamefont {Sj{\"{o}}blom}},\
  }\bibfield  {title} {\enquote {\bibinfo {title} {{Chemical destabilization of
  crude oil emulsions: Effect of nonionic surfactants as emulsion
  inhibitors}},}\ }\href {https://doi.org/10.1021/ef900355d} {\bibfield
  {journal} {\bibinfo  {journal} {Energy and Fuels}\ }\textbf {\bibinfo
  {volume} {23}},\ \bibinfo {pages} {4575--4583} (\bibinfo {year}
  {2009})}\BibitemShut {NoStop}%
\bibitem [{\citenamefont {Mirvakili}, \citenamefont {Rahimpour},\ and\
  \citenamefont {Jahanmiri}(2012)}]{Mirvakili2012}%
  \BibitemOpen
  \bibfield  {author} {\bibinfo {author} {\bibfnamefont {A.}~\bibnamefont
  {Mirvakili}}, \bibinfo {author} {\bibfnamefont {M.~R.}\ \bibnamefont
  {Rahimpour}},\ and\ \bibinfo {author} {\bibfnamefont {A.}~\bibnamefont
  {Jahanmiri}},\ }\bibfield  {title} {\enquote {\bibinfo {title} {{Effect of a
  cationic surfactant as a chemical destabilization of crude oil based
  emulsions and asphaltene stabilized}},}\ }\href
  {https://doi.org/10.1021/je2013268} {\bibfield  {journal} {\bibinfo
  {journal} {Journal of Chemical and Engineering Data}\ }\textbf {\bibinfo
  {volume} {57}},\ \bibinfo {pages} {1689--1699} (\bibinfo {year}
  {2012})}\BibitemShut {NoStop}%
\bibitem [{\citenamefont {Jia}\ \emph {et~al.}(2017)\citenamefont {Jia},
  \citenamefont {Leng}, \citenamefont {Hu}, \citenamefont {Song}, \citenamefont
  {Wu}, \citenamefont {Lian}, \citenamefont {Liang}, \citenamefont {Zhu},
  \citenamefont {Liu},\ and\ \citenamefont {Zhou}}]{Jia2017}%
  \BibitemOpen
  \bibfield  {author} {\bibinfo {author} {\bibfnamefont {H.}~\bibnamefont
  {Jia}}, \bibinfo {author} {\bibfnamefont {X.}~\bibnamefont {Leng}}, \bibinfo
  {author} {\bibfnamefont {M.}~\bibnamefont {Hu}}, \bibinfo {author}
  {\bibfnamefont {Y.}~\bibnamefont {Song}}, \bibinfo {author} {\bibfnamefont
  {H.}~\bibnamefont {Wu}}, \bibinfo {author} {\bibfnamefont {P.}~\bibnamefont
  {Lian}}, \bibinfo {author} {\bibfnamefont {Y.}~\bibnamefont {Liang}},
  \bibinfo {author} {\bibfnamefont {Y.}~\bibnamefont {Zhu}}, \bibinfo {author}
  {\bibfnamefont {J.}~\bibnamefont {Liu}},\ and\ \bibinfo {author}
  {\bibfnamefont {H.}~\bibnamefont {Zhou}},\ }\bibfield  {title} {\enquote
  {\bibinfo {title} {{Systematic investigation of the effects of mixed
  cationic/anionic surfactants on the interfacial tension of a water/model oil
  system and their application to enhance crude oil recovery}},}\ }\href
  {https://doi.org/10.1016/j.colsurfa.2017.06.055} {\bibfield  {journal}
  {\bibinfo  {journal} {Colloids and Surfaces A: Physicochemical and
  Engineering Aspects}\ }\textbf {\bibinfo {volume} {529}},\ \bibinfo {pages}
  {621--627} (\bibinfo {year} {2017})}\BibitemShut {NoStop}%
\bibitem [{\citenamefont {Kumari}\ \emph {et~al.}(2019)\citenamefont {Kumari},
  \citenamefont {Kakati}, \citenamefont {Nagarajan},\ and\ \citenamefont
  {Sangwai}}]{Kumari2019}%
  \BibitemOpen
  \bibfield  {author} {\bibinfo {author} {\bibfnamefont {R.}~\bibnamefont
  {Kumari}}, \bibinfo {author} {\bibfnamefont {A.}~\bibnamefont {Kakati}},
  \bibinfo {author} {\bibfnamefont {R.}~\bibnamefont {Nagarajan}},\ and\
  \bibinfo {author} {\bibfnamefont {J.~S.}\ \bibnamefont {Sangwai}},\
  }\bibfield  {title} {\enquote {\bibinfo {title} {{Synergistic effect of mixed
  anionic and cationic surfactant systems on the interfacial tension of crude
  oil-water and enhanced oil recovery}},}\ }\href
  {https://doi.org/10.1080/01932691.2018.1489280} {\bibfield  {journal}
  {\bibinfo  {journal} {Journal of Dispersion Science and Technology}\ }\textbf
  {\bibinfo {volume} {40}},\ \bibinfo {pages} {969--981} (\bibinfo {year}
  {2019})}\BibitemShut {NoStop}%
\bibitem [{\citenamefont {Nasiri}\ and\ \citenamefont
  {Biria}(2020)}]{Nasiri2020}%
  \BibitemOpen
  \bibfield  {author} {\bibinfo {author} {\bibfnamefont {M.~A.}\ \bibnamefont
  {Nasiri}}\ and\ \bibinfo {author} {\bibfnamefont {D.}~\bibnamefont {Biria}},\
  }\bibfield  {title} {\enquote {\bibinfo {title} {{Extraction of the
  indigenous crude oil dissolved biosurfactants and their potential in enhanced
  oil recovery}},}\ }\href {https://doi.org/10.1016/j.colsurfa.2020.125216}
  {\bibfield  {journal} {\bibinfo  {journal} {Colloids and Surfaces A:
  Physicochemical and Engineering Aspects}\ }\textbf {\bibinfo {volume}
  {603}},\ \bibinfo {pages} {125216} (\bibinfo {year} {2020})}\BibitemShut
  {NoStop}%
\bibitem [{\citenamefont {Samak}\ \emph {et~al.}(2020)\citenamefont {Samak},
  \citenamefont {Mahmoud}, \citenamefont {Aboulrous}, \citenamefont
  {Abdelhamid},\ and\ \citenamefont {Xing}}]{Samak2020}%
  \BibitemOpen
  \bibfield  {author} {\bibinfo {author} {\bibfnamefont {N.~A.}\ \bibnamefont
  {Samak}}, \bibinfo {author} {\bibfnamefont {T.}~\bibnamefont {Mahmoud}},
  \bibinfo {author} {\bibfnamefont {A.~A.}\ \bibnamefont {Aboulrous}}, \bibinfo
  {author} {\bibfnamefont {M.~M.}\ \bibnamefont {Abdelhamid}},\ and\ \bibinfo
  {author} {\bibfnamefont {J.}~\bibnamefont {Xing}},\ }\bibfield  {title}
  {\enquote {\bibinfo {title} {{Enhanced Biosurfactant Production Using
  Developed Fed-Batch Fermentation for Effective Heavy Crude Oil Recovery}},}\
  }\href {https://doi.org/10.1021/acs.energyfuels.0c02676} {\bibfield
  {journal} {\bibinfo  {journal} {Energy and Fuels}\ }\textbf {\bibinfo
  {volume} {34}},\ \bibinfo {pages} {14560--14572} (\bibinfo {year}
  {2020})}\BibitemShut {NoStop}%
\bibitem [{\citenamefont {Yang}\ \emph {et~al.}(2020)\citenamefont {Yang},
  \citenamefont {Zu}, \citenamefont {Zhu}, \citenamefont {Jin}, \citenamefont
  {Cui},\ and\ \citenamefont {Long}}]{Yang2020}%
  \BibitemOpen
  \bibfield  {author} {\bibinfo {author} {\bibfnamefont {Z.}~\bibnamefont
  {Yang}}, \bibinfo {author} {\bibfnamefont {Y.}~\bibnamefont {Zu}}, \bibinfo
  {author} {\bibfnamefont {J.}~\bibnamefont {Zhu}}, \bibinfo {author}
  {\bibfnamefont {M.}~\bibnamefont {Jin}}, \bibinfo {author} {\bibfnamefont
  {T.}~\bibnamefont {Cui}},\ and\ \bibinfo {author} {\bibfnamefont
  {X.}~\bibnamefont {Long}},\ }\bibfield  {title} {\enquote {\bibinfo {title}
  {{Application of biosurfactant surfactin as a pH-switchable biodemulsifier
  for efficient oil recovery from waste crude oil}},}\ }\href
  {https://doi.org/10.1016/j.chemosphere.2019.124946} {\bibfield  {journal}
  {\bibinfo  {journal} {Chemosphere}\ }\textbf {\bibinfo {volume} {240}},\
  \bibinfo {pages} {1--8} (\bibinfo {year} {2020})}\BibitemShut {NoStop}%
\bibitem [{\citenamefont {Sakthivel}\ \emph {et~al.}(2017)\citenamefont
  {Sakthivel}, \citenamefont {Velusamy}, \citenamefont {Nair}, \citenamefont
  {Sharma},\ and\ \citenamefont {Sangwai}}]{Sakthivel2017}%
  \BibitemOpen
  \bibfield  {author} {\bibinfo {author} {\bibfnamefont {S.}~\bibnamefont
  {Sakthivel}}, \bibinfo {author} {\bibfnamefont {S.}~\bibnamefont {Velusamy}},
  \bibinfo {author} {\bibfnamefont {V.~C.}\ \bibnamefont {Nair}}, \bibinfo
  {author} {\bibfnamefont {T.}~\bibnamefont {Sharma}},\ and\ \bibinfo {author}
  {\bibfnamefont {J.~S.}\ \bibnamefont {Sangwai}},\ }\bibfield  {title}
  {\enquote {\bibinfo {title} {{Interfacial tension of crude oil-water system
  with imidazolium and lactam-based ionic liquids and their evaluation for
  enhanced oil recovery under high saline environment}},}\ }\href
  {https://doi.org/10.1016/j.fuel.2016.11.064} {\bibfield  {journal} {\bibinfo
  {journal} {Fuel}\ }\textbf {\bibinfo {volume} {191}},\ \bibinfo {pages}
  {239--250} (\bibinfo {year} {2017})}\BibitemShut {NoStop}%
\bibitem [{\citenamefont {Jia}\ \emph {et~al.}(2018)\citenamefont {Jia},
  \citenamefont {Lian}, \citenamefont {Liang}, \citenamefont {Zhu},
  \citenamefont {Huang}, \citenamefont {Wu}, \citenamefont {Leng},\ and\
  \citenamefont {Zhou}}]{Jia2018}%
  \BibitemOpen
  \bibfield  {author} {\bibinfo {author} {\bibfnamefont {H.}~\bibnamefont
  {Jia}}, \bibinfo {author} {\bibfnamefont {P.}~\bibnamefont {Lian}}, \bibinfo
  {author} {\bibfnamefont {Y.}~\bibnamefont {Liang}}, \bibinfo {author}
  {\bibfnamefont {Y.}~\bibnamefont {Zhu}}, \bibinfo {author} {\bibfnamefont
  {P.}~\bibnamefont {Huang}}, \bibinfo {author} {\bibfnamefont
  {H.}~\bibnamefont {Wu}}, \bibinfo {author} {\bibfnamefont {X.}~\bibnamefont
  {Leng}},\ and\ \bibinfo {author} {\bibfnamefont {H.}~\bibnamefont {Zhou}},\
  }\bibfield  {title} {\enquote {\bibinfo {title} {{Systematic Investigation of
  the Effects of Zwitterionic Surface-Active Ionic Liquids on the Interfacial
  Tension of a Water/Crude Oil System and Their Application to Enhance Crude
  Oil Recovery}},}\ }\href {https://doi.org/10.1021/acs.energyfuels.7b02746}
  {\bibfield  {journal} {\bibinfo  {journal} {Energy and Fuels}\ }\textbf
  {\bibinfo {volume} {32}},\ \bibinfo {pages} {154--160} (\bibinfo {year}
  {2018})}\BibitemShut {NoStop}%
\bibitem [{\citenamefont {Liu}\ \emph {et~al.}(2019)\citenamefont {Liu},
  \citenamefont {Huang}, \citenamefont {Feng}, \citenamefont {Lian},
  \citenamefont {Liang}, \citenamefont {Huang}, \citenamefont {Yan},\ and\
  \citenamefont {Jia}}]{Liu2019b}%
  \BibitemOpen
  \bibfield  {author} {\bibinfo {author} {\bibfnamefont {J.}~\bibnamefont
  {Liu}}, \bibinfo {author} {\bibfnamefont {P.}~\bibnamefont {Huang}}, \bibinfo
  {author} {\bibfnamefont {Q.}~\bibnamefont {Feng}}, \bibinfo {author}
  {\bibfnamefont {P.}~\bibnamefont {Lian}}, \bibinfo {author} {\bibfnamefont
  {Y.}~\bibnamefont {Liang}}, \bibinfo {author} {\bibfnamefont
  {W.}~\bibnamefont {Huang}}, \bibinfo {author} {\bibfnamefont
  {H.}~\bibnamefont {Yan}},\ and\ \bibinfo {author} {\bibfnamefont
  {H.}~\bibnamefont {Jia}},\ }\bibfield  {title} {\enquote {\bibinfo {title}
  {{Systematic investigation of the effects of an anionic surface active ionic
  liquid on the interfacial tension of a water/crude oil system and its
  application to enhance crude oil recovery}},}\ }\href
  {https://doi.org/10.1080/01932691.2018.1527230} {\bibfield  {journal}
  {\bibinfo  {journal} {Journal of Dispersion Science and Technology}\ }\textbf
  {\bibinfo {volume} {40}},\ \bibinfo {pages} {1657--1663} (\bibinfo {year}
  {2019})}\BibitemShut {NoStop}%
\bibitem [{\citenamefont {Chen}\ \emph {et~al.}(2017)\citenamefont {Chen},
  \citenamefont {Kaufman}, \citenamefont {Kristiansen}, \citenamefont {Seo},
  \citenamefont {Schrader}, \citenamefont {Alotaibi}, \citenamefont {Dobbs},
  \citenamefont {Cadirov}, \citenamefont {Boles}, \citenamefont {Ayirala},
  \citenamefont {Israelachvili},\ and\ \citenamefont {Yousef}}]{Chen2017}%
  \BibitemOpen
  \bibfield  {author} {\bibinfo {author} {\bibfnamefont {S.~Y.}\ \bibnamefont
  {Chen}}, \bibinfo {author} {\bibfnamefont {Y.}~\bibnamefont {Kaufman}},
  \bibinfo {author} {\bibfnamefont {K.}~\bibnamefont {Kristiansen}}, \bibinfo
  {author} {\bibfnamefont {D.}~\bibnamefont {Seo}}, \bibinfo {author}
  {\bibfnamefont {A.~M.}\ \bibnamefont {Schrader}}, \bibinfo {author}
  {\bibfnamefont {M.~B.}\ \bibnamefont {Alotaibi}}, \bibinfo {author}
  {\bibfnamefont {H.~A.}\ \bibnamefont {Dobbs}}, \bibinfo {author}
  {\bibfnamefont {N.~A.}\ \bibnamefont {Cadirov}}, \bibinfo {author}
  {\bibfnamefont {J.~R.}\ \bibnamefont {Boles}}, \bibinfo {author}
  {\bibfnamefont {S.~C.}\ \bibnamefont {Ayirala}}, \bibinfo {author}
  {\bibfnamefont {J.~N.}\ \bibnamefont {Israelachvili}},\ and\ \bibinfo
  {author} {\bibfnamefont {A.~A.}\ \bibnamefont {Yousef}},\ }\bibfield  {title}
  {\enquote {\bibinfo {title} {{Effects of Salinity on Oil Recovery (the
  "dilution Effect"): Experimental and Theoretical Studies of Crude
  Oil/Brine/Carbonate Surface Restructuring and Associated Physicochemical
  Interactions}},}\ }\href {https://doi.org/10.1021/acs.energyfuels.7b00869}
  {\bibfield  {journal} {\bibinfo  {journal} {Energy and Fuels}\ }\textbf
  {\bibinfo {volume} {31}},\ \bibinfo {pages} {8925--8941} (\bibinfo {year}
  {2017})}\BibitemShut {NoStop}%
\bibitem [{\citenamefont {Bidhendi}\ \emph {et~al.}(2018)\citenamefont
  {Bidhendi}, \citenamefont {Garcia-Olvera}, \citenamefont {Morin},
  \citenamefont {Oakey},\ and\ \citenamefont {Alvarado}}]{Bidhendi2018}%
  \BibitemOpen
  \bibfield  {author} {\bibinfo {author} {\bibfnamefont {M.~M.}\ \bibnamefont
  {Bidhendi}}, \bibinfo {author} {\bibfnamefont {G.}~\bibnamefont
  {Garcia-Olvera}}, \bibinfo {author} {\bibfnamefont {B.}~\bibnamefont
  {Morin}}, \bibinfo {author} {\bibfnamefont {J.~S.}\ \bibnamefont {Oakey}},\
  and\ \bibinfo {author} {\bibfnamefont {V.}~\bibnamefont {Alvarado}},\
  }\bibfield  {title} {\enquote {\bibinfo {title} {{Interfacial Viscoelasticity
  of Crude Oil/Brine: An Alternative Enhanced-Oil-Recovery Mechanism in Smart
  Waterflooding}},}\ }\href@noop {} {\bibfield  {journal} {\bibinfo  {journal}
  {SPE Journal}\ }\textbf {\bibinfo {volume} {23}},\ \bibinfo {pages}
  {803--818} (\bibinfo {year} {2018})}\BibitemShut {NoStop}%
\bibitem [{\citenamefont {Takeya}\ \emph {et~al.}(2019)\citenamefont {Takeya},
  \citenamefont {Shimokawara}, \citenamefont {Elakneswaran}, \citenamefont
  {Nawa},\ and\ \citenamefont {Takahashi}}]{Takeya2019}%
  \BibitemOpen
  \bibfield  {author} {\bibinfo {author} {\bibfnamefont {M.}~\bibnamefont
  {Takeya}}, \bibinfo {author} {\bibfnamefont {M.}~\bibnamefont {Shimokawara}},
  \bibinfo {author} {\bibfnamefont {Y.}~\bibnamefont {Elakneswaran}}, \bibinfo
  {author} {\bibfnamefont {T.}~\bibnamefont {Nawa}},\ and\ \bibinfo {author}
  {\bibfnamefont {S.}~\bibnamefont {Takahashi}},\ }\bibfield  {title} {\enquote
  {\bibinfo {title} {{Predicting the electrokinetic properties of the crude
  oil/brine interface for enhanced oil recovery in low salinity water
  flooding}},}\ }\href {https://doi.org/10.1016/j.fuel.2018.08.079} {\bibfield
  {journal} {\bibinfo  {journal} {Fuel}\ }\textbf {\bibinfo {volume} {235}},\
  \bibinfo {pages} {822--831} (\bibinfo {year} {2019})}\BibitemShut {NoStop}%
\bibitem [{\citenamefont {Sanyal}\ \emph {et~al.}(2019)\citenamefont {Sanyal},
  \citenamefont {Bhui}, \citenamefont {Balaga},\ and\ \citenamefont
  {Kumar}}]{Sanyal2019}%
  \BibitemOpen
  \bibfield  {author} {\bibinfo {author} {\bibfnamefont {S.}~\bibnamefont
  {Sanyal}}, \bibinfo {author} {\bibfnamefont {U.~K.}\ \bibnamefont {Bhui}},
  \bibinfo {author} {\bibfnamefont {D.}~\bibnamefont {Balaga}},\ and\ \bibinfo
  {author} {\bibfnamefont {S.~S.}\ \bibnamefont {Kumar}},\ }\bibfield  {title}
  {\enquote {\bibinfo {title} {{Interaction study of montmorillonite-crude
  oil-brine: Molecular-level implications on enhanced oil recovery during low
  saline water flooding from hydrocarbon reservoirs}},}\ }\href
  {https://doi.org/10.1016/j.fuel.2019.115725} {\bibfield  {journal} {\bibinfo
  {journal} {Fuel}\ }\textbf {\bibinfo {volume} {254}},\ \bibinfo {pages}
  {115725} (\bibinfo {year} {2019})}\BibitemShut {NoStop}%
\bibitem [{\citenamefont {{Ilia Anisa}}, \citenamefont {Nour},\ and\
  \citenamefont {Nour}(2011)}]{IliaAnisa2011}%
  \BibitemOpen
  \bibfield  {author} {\bibinfo {author} {\bibfnamefont {A.~N.}\ \bibnamefont
  {{Ilia Anisa}}}, \bibinfo {author} {\bibfnamefont {A.~H.}\ \bibnamefont
  {Nour}},\ and\ \bibinfo {author} {\bibfnamefont {A.~H.}\ \bibnamefont
  {Nour}},\ }\href {https://doi.org/10.3923/jas.2011.2898.2906} {\enquote
  {\bibinfo {title} {{Destabilization of heavy and light crude oil emulsions
  via microwave heating technology: An optimization study}},}\ } (\bibinfo
  {year} {2011})\BibitemShut {NoStop}%
\bibitem [{\citenamefont {Rodionova}, \citenamefont {KeleÅŸoÇ§lu},\ and\
  \citenamefont {Sj{\"{o}}blom}(2014)}]{Rodionova2014}%
  \BibitemOpen
  \bibfield  {author} {\bibinfo {author} {\bibfnamefont {G.}~\bibnamefont
  {Rodionova}}, \bibinfo {author} {\bibfnamefont {S.}~\bibnamefont
  {KeleÅŸoÇ§lu}},\ and\ \bibinfo {author} {\bibfnamefont {J.}~\bibnamefont
  {Sj{\"{o}}blom}},\ }\bibfield  {title} {\enquote {\bibinfo {title} {{AC field
  induced destabilization of water-in-oil emulsions based on North Sea acidic
  crude oil}},}\ }\href {https://doi.org/10.1016/j.colsurfa.2014.01.019}
  {\bibfield  {journal} {\bibinfo  {journal} {Colloids and Surfaces A:
  Physicochemical and Engineering Aspects}\ }\textbf {\bibinfo {volume}
  {448}},\ \bibinfo {pages} {60--66} (\bibinfo {year} {2014})}\BibitemShut
  {NoStop}%
\bibitem [{\citenamefont {Bremond}, \citenamefont {Dom{\'{e}}jean},\ and\
  \citenamefont {Bibette}(2011)}]{Bremond2011}%
  \BibitemOpen
  \bibfield  {author} {\bibinfo {author} {\bibfnamefont {N.}~\bibnamefont
  {Bremond}}, \bibinfo {author} {\bibfnamefont {H.}~\bibnamefont
  {Dom{\'{e}}jean}},\ and\ \bibinfo {author} {\bibfnamefont {J.}~\bibnamefont
  {Bibette}},\ }\bibfield  {title} {\enquote {\bibinfo {title} {{Propagation of
  drop coalescence in a two-dimensional emulsion: A route towards phase
  inversion}},}\ }\href {https://doi.org/10.1103/PhysRevLett.106.214502}
  {\bibfield  {journal} {\bibinfo  {journal} {Physical Review Letters}\
  }\textbf {\bibinfo {volume} {106}},\ \bibinfo {pages} {1--4} (\bibinfo {year}
  {2011})}\BibitemShut {NoStop}%
\bibitem [{\citenamefont {Brailsford}\ and\ \citenamefont
  {Wynblatt}(1979)}]{Brailsford1979}%
  \BibitemOpen
  \bibfield  {author} {\bibinfo {author} {\bibfnamefont {A.~D.}\ \bibnamefont
  {Brailsford}}\ and\ \bibinfo {author} {\bibfnamefont {P.}~\bibnamefont
  {Wynblatt}},\ }\bibfield  {title} {\enquote {\bibinfo {title} {{The
  dependence of Ostwald ripening kinetics on particle volume fraction}},}\
  }\href {https://doi.org/10.1016/0001-6160(79)90041-5} {\bibfield  {journal}
  {\bibinfo  {journal} {Acta Metallurgica}\ }\textbf {\bibinfo {volume} {27}},\
  \bibinfo {pages} {489--497} (\bibinfo {year} {1979})}\BibitemShut {NoStop}%
\bibitem [{\citenamefont {Yao}\ \emph {et~al.}(1992)\citenamefont {Yao},
  \citenamefont {Elder}, \citenamefont {Guo},\ and\ \citenamefont
  {Grant}}]{Yao1992}%
  \BibitemOpen
  \bibfield  {author} {\bibinfo {author} {\bibfnamefont {J.~H.}\ \bibnamefont
  {Yao}}, \bibinfo {author} {\bibfnamefont {K.~R.}\ \bibnamefont {Elder}},
  \bibinfo {author} {\bibfnamefont {H.}~\bibnamefont {Guo}},\ and\ \bibinfo
  {author} {\bibfnamefont {M.}~\bibnamefont {Grant}},\ }\bibfield  {title}
  {\enquote {\bibinfo {title} {{Ostwald ripening in two and three
  dimensions}},}\ }\href {https://doi.org/10.1103/PhysRevB.45.8173} {\bibfield
  {journal} {\bibinfo  {journal} {Physical Review B}\ }\textbf {\bibinfo
  {volume} {45}},\ \bibinfo {pages} {8173--8176} (\bibinfo {year}
  {1992})}\BibitemShut {NoStop}%
\bibitem [{\citenamefont {Deblais}\ \emph {et~al.}(2015)\citenamefont
  {Deblais}, \citenamefont {Harich}, \citenamefont {Bonn}, \citenamefont
  {Colin},\ and\ \citenamefont {Kellay}}]{Deblais2015}%
  \BibitemOpen
  \bibfield  {author} {\bibinfo {author} {\bibfnamefont {A.}~\bibnamefont
  {Deblais}}, \bibinfo {author} {\bibfnamefont {R.}~\bibnamefont {Harich}},
  \bibinfo {author} {\bibfnamefont {D.}~\bibnamefont {Bonn}}, \bibinfo {author}
  {\bibfnamefont {A.}~\bibnamefont {Colin}},\ and\ \bibinfo {author}
  {\bibfnamefont {H.}~\bibnamefont {Kellay}},\ }\bibfield  {title} {\enquote
  {\bibinfo {title} {{Spreading of an Oil-in-Water Emulsion on a Glass Plate:
  Phase Inversion and Pattern Formation}},}\ }\href
  {https://doi.org/10.1021/la504639q} {\bibfield  {journal} {\bibinfo
  {journal} {Langmuir}\ }\textbf {\bibinfo {volume} {31}},\ \bibinfo {pages}
  {5971--5981} (\bibinfo {year} {2015})}\BibitemShut {NoStop}%
\bibitem [{\citenamefont {Dekker}\ \emph {et~al.}(2020)\citenamefont {Dekker},
  \citenamefont {Deblais}, \citenamefont {Velikov}, \citenamefont {Veenstra},
  \citenamefont {Colin}, \citenamefont {Kellay}, \citenamefont {Kegel},\ and\
  \citenamefont {Bonn}}]{Dekker2020}%
  \BibitemOpen
  \bibfield  {author} {\bibinfo {author} {\bibfnamefont {R.~I.}\ \bibnamefont
  {Dekker}}, \bibinfo {author} {\bibfnamefont {A.}~\bibnamefont {Deblais}},
  \bibinfo {author} {\bibfnamefont {K.~P.}\ \bibnamefont {Velikov}}, \bibinfo
  {author} {\bibfnamefont {P.}~\bibnamefont {Veenstra}}, \bibinfo {author}
  {\bibfnamefont {A.}~\bibnamefont {Colin}}, \bibinfo {author} {\bibfnamefont
  {H.}~\bibnamefont {Kellay}}, \bibinfo {author} {\bibfnamefont {W.~K.}\
  \bibnamefont {Kegel}},\ and\ \bibinfo {author} {\bibfnamefont
  {D.}~\bibnamefont {Bonn}},\ }\bibfield  {title} {\enquote {\bibinfo {title}
  {{Emulsion Destabilization by Squeeze Flow}},}\ }\href
  {https://doi.org/10.1021/acs.langmuir.0c00759} {\bibfield  {journal}
  {\bibinfo  {journal} {Langmuir}\ }\textbf {\bibinfo {volume} {36}},\ \bibinfo
  {pages} {7795--7800} (\bibinfo {year} {2020})}\BibitemShut {NoStop}%
\bibitem [{\citenamefont {Hartland}\ and\ \citenamefont
  {Jeelani}(1987)}]{Hartland1987}%
  \BibitemOpen
  \bibfield  {author} {\bibinfo {author} {\bibfnamefont {S.}~\bibnamefont
  {Hartland}}\ and\ \bibinfo {author} {\bibfnamefont {S.}~\bibnamefont
  {Jeelani}},\ }\bibfield  {title} {\enquote {\bibinfo {title} {{Choice of
  model for predicting the dispersion height in liquid/liquid gravity settlers
  from batch settling data}},}\ }\href@noop {} {\bibfield  {journal} {\bibinfo
  {journal} {Chemical Engineering Science}\ }\textbf {\bibinfo {volume} {42}},\
  \bibinfo {pages} {1927--1938} (\bibinfo {year} {1987})}\BibitemShut {NoStop}%
\bibitem [{\citenamefont {Aleem}\ \emph {et~al.}(2020)\citenamefont {Aleem},
  \citenamefont {Mellon}, \citenamefont {Khan},\ and\ \citenamefont
  {Al-Kayiem}}]{Aleem2020}%
  \BibitemOpen
  \bibfield  {author} {\bibinfo {author} {\bibfnamefont {W.}~\bibnamefont
  {Aleem}}, \bibinfo {author} {\bibfnamefont {N.}~\bibnamefont {Mellon}},
  \bibinfo {author} {\bibfnamefont {J.~A.}\ \bibnamefont {Khan}},\ and\
  \bibinfo {author} {\bibfnamefont {H.~H.}\ \bibnamefont {Al-Kayiem}},\
  }\bibfield  {title} {\enquote {\bibinfo {title} {{Experimental investigation
  and mathematical modeling of oil/water emulsion separation effectiveness
  containing alkali-surfactant-polymer}},}\ }\href
  {https://doi.org/10.1080/01932691.2020.1738244} {\bibfield  {journal}
  {\bibinfo  {journal} {Journal of Dispersion Science and Technology}\ }\textbf
  {\bibinfo {volume} {0}},\ \bibinfo {pages} {1--13} (\bibinfo {year}
  {2020})}\BibitemShut {NoStop}%
\bibitem [{\citenamefont {Paredes}, \citenamefont {Shahidzadeh},\ and\
  \citenamefont {Bonn}(2015)}]{Paredes2015}%
  \BibitemOpen
  \bibfield  {author} {\bibinfo {author} {\bibfnamefont {J.}~\bibnamefont
  {Paredes}}, \bibinfo {author} {\bibfnamefont {N.}~\bibnamefont
  {Shahidzadeh}},\ and\ \bibinfo {author} {\bibfnamefont {D.}~\bibnamefont
  {Bonn}},\ }\bibfield  {title} {\enquote {\bibinfo {title} {{Wall slip and
  fluidity in emulsion flow}},}\ }\href
  {https://doi.org/10.1103/PhysRevE.92.042313} {\bibfield  {journal} {\bibinfo
  {journal} {Physical Review E - Statistical, Nonlinear, and Soft Matter
  Physics}\ }\textbf {\bibinfo {volume} {92}},\ \bibinfo {pages} {1--7}
  (\bibinfo {year} {2015})}\BibitemShut {NoStop}%
\bibitem [{\citenamefont {Paredes}, \citenamefont {Michels},\ and\
  \citenamefont {Bonn}(2013)}]{Paredes2013}%
  \BibitemOpen
  \bibfield  {author} {\bibinfo {author} {\bibfnamefont {J.}~\bibnamefont
  {Paredes}}, \bibinfo {author} {\bibfnamefont {M.~A.}\ \bibnamefont
  {Michels}},\ and\ \bibinfo {author} {\bibfnamefont {D.}~\bibnamefont
  {Bonn}},\ }\bibfield  {title} {\enquote {\bibinfo {title} {{Rheology across
  the zero-temperature jamming transition}},}\ }\href
  {https://doi.org/10.1103/PhysRevLett.111.015701} {\bibfield  {journal}
  {\bibinfo  {journal} {Physical Review Letters}\ }\textbf {\bibinfo {volume}
  {111}},\ \bibinfo {pages} {1--5} (\bibinfo {year} {2013})}\BibitemShut
  {NoStop}%
\bibitem [{\citenamefont {{De Cagny}}\ \emph {et~al.}(2016)\citenamefont {{De
  Cagny}}, \citenamefont {Vos}, \citenamefont {Vahabi}, \citenamefont
  {Kurniawan}, \citenamefont {Doi}, \citenamefont {Koenderink}, \citenamefont
  {MacKintosh},\ and\ \citenamefont {Bonn}}]{DeCagny2016}%
  \BibitemOpen
  \bibfield  {author} {\bibinfo {author} {\bibfnamefont {H.~C.}\ \bibnamefont
  {{De Cagny}}}, \bibinfo {author} {\bibfnamefont {B.~E.}\ \bibnamefont {Vos}},
  \bibinfo {author} {\bibfnamefont {M.}~\bibnamefont {Vahabi}}, \bibinfo
  {author} {\bibfnamefont {N.~A.}\ \bibnamefont {Kurniawan}}, \bibinfo {author}
  {\bibfnamefont {M.}~\bibnamefont {Doi}}, \bibinfo {author} {\bibfnamefont
  {G.~H.}\ \bibnamefont {Koenderink}}, \bibinfo {author} {\bibfnamefont
  {F.~C.}\ \bibnamefont {MacKintosh}},\ and\ \bibinfo {author} {\bibfnamefont
  {D.}~\bibnamefont {Bonn}},\ }\bibfield  {title} {\enquote {\bibinfo {title}
  {{Porosity governs normal stresses in polymer gels}},}\ }\href
  {https://doi.org/10.1103/PhysRevLett.117.217802} {\bibfield  {journal}
  {\bibinfo  {journal} {Physical Review Letters}\ }\textbf {\bibinfo {volume}
  {117}},\ \bibinfo {pages} {1--5} (\bibinfo {year} {2016})},\ \Eprint
  {https://arxiv.org/abs/1605.04139} {arXiv:1605.04139} \BibitemShut {NoStop}%
\bibitem [{\citenamefont {Burgers}(1935)}]{Burgers1935}%
  \BibitemOpen
  \bibfield  {author} {\bibinfo {author} {\bibfnamefont {J.}~\bibnamefont
  {Burgers}},\ }\href@noop {} {\emph {\bibinfo {title} {{Mechanical
  considerations, model systems, phenomenological theories of relaxation and of
  viscosity, 1st report on viscosity and plasticity}}}}\ (\bibinfo  {publisher}
  {Nordemann, New York},\ \bibinfo {year} {1935})\BibitemShut {NoStop}%
\bibitem [{\citenamefont {Mezger}(2011)}]{Mezger2011}%
  \BibitemOpen
  \bibfield  {author} {\bibinfo {author} {\bibfnamefont {T.~G.}\ \bibnamefont
  {Mezger}},\ }\href@noop {} {\emph {\bibinfo {title} {{The Rheology
  Handbook}}}},\ \bibinfo {edition} {3rd}\ ed.\ (\bibinfo  {publisher}
  {Vincentz Network},\ \bibinfo {year} {2011})\ p.\ \bibinfo {pages}
  {434}\BibitemShut {NoStop}%
\bibitem [{\citenamefont {Schmalholz}\ and\ \citenamefont
  {Podladchikov}(2001)}]{Schmalholz2001}%
  \BibitemOpen
  \bibfield  {author} {\bibinfo {author} {\bibfnamefont {S.}~\bibnamefont
  {Schmalholz}}\ and\ \bibinfo {author} {\bibfnamefont {Y.}~\bibnamefont
  {Podladchikov}},\ }\bibfield  {title} {\enquote {\bibinfo {title}
  {{Viscoelastic Folding: Maxwell versus Kelvin Rheology}},}\ }\href@noop {}
  {\bibfield  {journal} {\bibinfo  {journal} {Geophysical Research Letters}\
  }\textbf {\bibinfo {volume} {28}},\ \bibinfo {pages} {1835--1838} (\bibinfo
  {year} {2001})}\BibitemShut {NoStop}%
\bibitem [{\citenamefont {Chatzigiannakis}\ \emph {et~al.}(2020)\citenamefont
  {Chatzigiannakis}, \citenamefont {Veenstra}, \citenamefont {{Ten Bosch}},\
  and\ \citenamefont {Vermant}}]{Chatzigiannakis2020}%
  \BibitemOpen
  \bibfield  {author} {\bibinfo {author} {\bibfnamefont {E.}~\bibnamefont
  {Chatzigiannakis}}, \bibinfo {author} {\bibfnamefont {P.}~\bibnamefont
  {Veenstra}}, \bibinfo {author} {\bibfnamefont {D.}~\bibnamefont {{Ten
  Bosch}}},\ and\ \bibinfo {author} {\bibfnamefont {J.}~\bibnamefont
  {Vermant}},\ }\bibfield  {title} {\enquote {\bibinfo {title} {{Mimicking
  coalescence using a pressure-controlled dynamic thin film balance}},}\ }\href
  {https://doi.org/10.1039/d0sm00784f} {\bibfield  {journal} {\bibinfo
  {journal} {Soft Matter}\ }\textbf {\bibinfo {volume} {16}},\ \bibinfo {pages}
  {9410--9422} (\bibinfo {year} {2020})}\BibitemShut {NoStop}%
\end{thebibliography}
\end{document}